\documentclass[conference,compsoc]{IEEEtran}
\ifCLASSOPTIONcompsoc
  \usepackage[nocompress]{cite}
\else
  \usepackage{cite}
\fi
\ifCLASSINFOpdf
\else
\fi
\usepackage{graphicx} 
\usepackage{hyperref} 
\usepackage{cleveref}
\usepackage{algorithmic}
\usepackage{graphicx}
\usepackage{textcomp}
\usepackage{xcolor}
\usepackage{multirow}
\usepackage{xspace}
\usepackage{listings}
\usepackage{caption}
\usepackage{xcolor}

\newcommand{\mysection}[1]{\vspace{-.07in}\section{#1}\vspace{-.12in}}

\begin{document}
%
\title{Exploring the Limits of ChatGPT in Software Security Applications}

\author{
  \qquad Fangzhou Wu$^{{1}\ast}$ \qquad Qingzhao Zhang$^{{2}\ast}$ \qquad Ati Priya Bajaj$^{2}$ \qquad  Tiffany Bao$^{3}$ \qquad \\ \medskip 
Ning Zhang$^{4}$ \qquad Ruoyu "Fish" Wang$^{3}$ \qquad 
Chaowei Xiao$^{{1}\dagger}$ \\ \medskip
 \small $^{1}$University of Wisconsin, Madison \qquad $^{2}$University of Michigan, Ann Arbor \qquad $^{3}$ASU \qquad $^{4}$Washington University in St. Louis 
}

\maketitle

\def\thefootnote{$\ast$}\footnotetext{Equal contributors}\def\thefootnote{\arabic{footnote}}
\def\thefootnote{$\dagger$}\footnotetext{Corresponding Author: cxiao34@wisc.edu}

\begin{abstract}
Large language models (LLMs) have undergone rapid evolution and achieved remarkable results in recent times. OpenAI's ChatGPT, backed by GPT-3.5 or GPT-4, has gained instant popularity due to its strong capability across a wide range of tasks, including natural language tasks, coding, mathematics, and engaging conversations. However, the impacts and limits of such LLMs in system security domain are less explored. 
In this paper, we delve into the limits of LLMs (i.e., ChatGPT)  in seven software security applications including vulnerability detection/repair, debugging, debloating, decompilation, patching, root cause analysis, symbolic execution, and fuzzing. Our exploration reveals that ChatGPT not only excels at generating code, which is the conventional application of language models, but also demonstrates strong capability in understanding user-provided commands in natural languages, reasoning about control and data flows within programs, generating complex data structures, and even decompiling assembly code. Notably, GPT-4 showcases significant improvements over GPT-3.5 in most security tasks. Also, certain limitations of ChatGPT in security-related tasks are identified, such as its constrained ability to process long code contexts. 
\end{abstract}


\mysection{Introduction}
\label{sec:introduction}

\begin{figure}[h]
    \centering
    \includegraphics[width=0.40\textwidth]{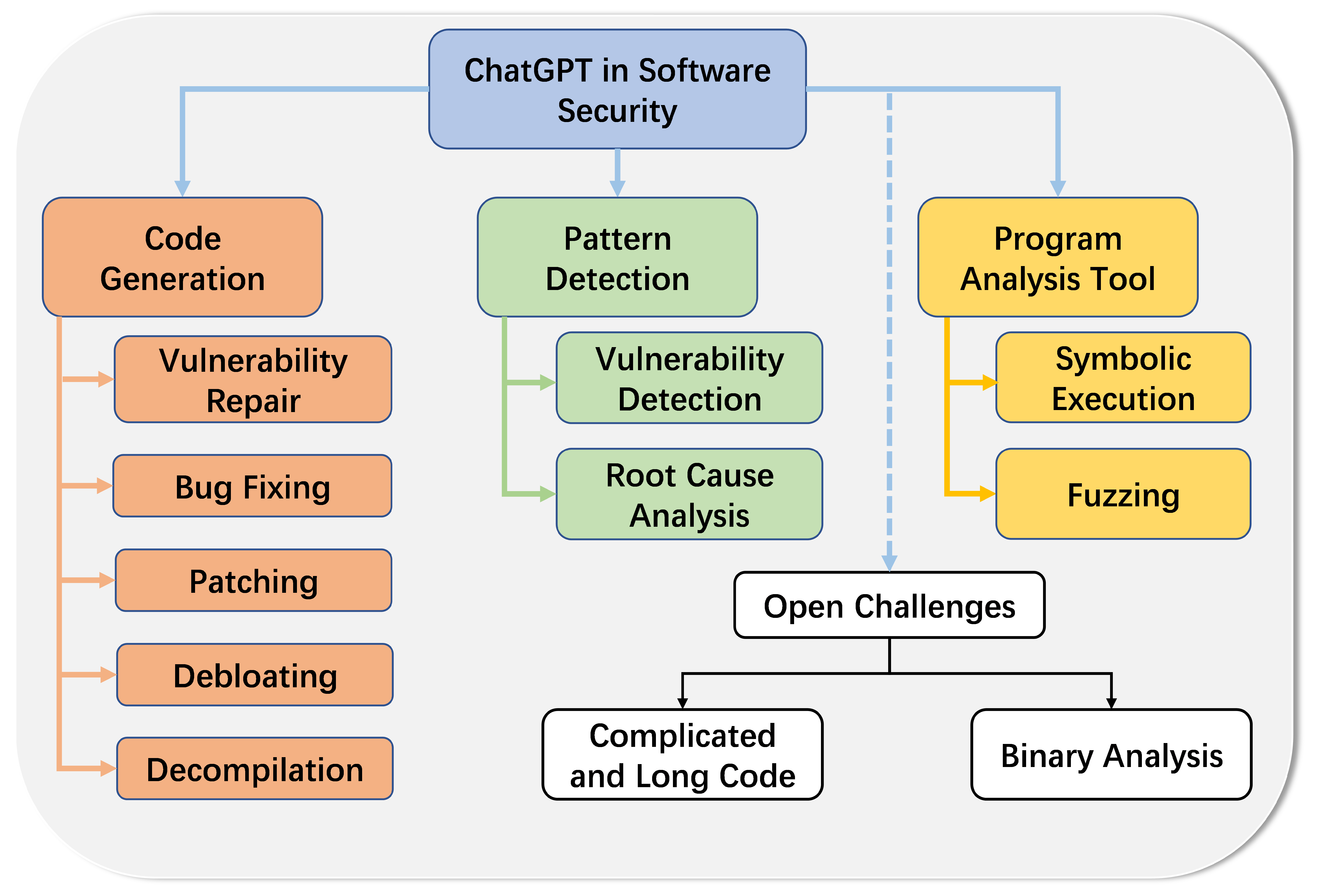}
    \caption{Categories of ChatGPT-based software security tasks and open research questions covered by this work.}
    \label{fig:category_chart}
\end{figure}
\vspace{0.05in}

The recent breakthrough of large language models (LLMs) is a revolution of the research of natural language processing (NLP).
Since the invention of neural networks back in 1980s~\cite{ rumelhart1986learning}, NLP techniques evolves along with the development of more complex model structures. 
From Recurrent Neural Networks (RNNs)~\cite{rumelhart1985learning,jordan1997serial} to Long Short-Term Memory networks (LSTMs)~\cite{hochreiter1997long}, the model structure is getting more diverse. 
In 2017, Vaswani et. al.~\cite{vaswani2017attention} proposed transformer models, which was a major breakthrough and opened the door to large complex models.
Since then, industry players have invested significant efforts in training transformer-based language models with a large number of model parameters, i.e., Large Language Models (LLMs). OpenAI's Generative Pretrained Transformer (GPT) models~\cite{radford2018improving,radford2019language,brown2020language,gpt-4} are typical examples of such powerful LLMs. GPT models are pre-trained auto-regressive models that can generate text.
From the initial GPT-1~\cite{radford2018improving} released in 2018 to the latest GPT-4~\cite{gpt-4} released in 2023, the number of model parameters grows from 0.12 billion to 1 trillion. 
Within the trend of LLMs in 2023, ChatGPT~\cite{chatgpt}, a chatbot backed by GPT-3.5 or GPT-4 models, is one of the most remarkable achievements. It gains public popularity instantly after its release with amazingly high-quality AI-generated conversations. A large body of research~\cite{aljanabi2023chatgpt,gozalo2023chatgpt,aljanabi2023chatgpt,liu2023summary,bubeck2023sparks,azaria2022chatgpt} conduct measurement on ChatGPT and they demonstrate the strong capability of ChatGPT in a variety of tasks. The studies demonstrate that ChatGPT is able to not only generate natural languages but also generate tables or figures for data visualization~\cite{maddigan2023chat2vis,bubeck2023sparks}, solve mathematical problems~\cite{bubeck2023sparks,frieder2023mathematical}, play games~\cite{biswas2023role,tsai2023can}, assist healthcare services~\cite{sallam2023chatgpt}, search for neural network architecture~\cite{zheng2023can}, pass real-world exams~\cite{choi2023chatgpt,ryznar2023exams,newton2023chatgpt}, and more~\cite{bang2023multitask}, though limitation exists~\cite{gozalo2023chatgpt,borji2023categorical}, e.g., ChatGPT may give wrong or improper answers to certain questions.
Bubeck et. al.~\cite{bubeck2023sparks} do case studies on GPT-4's problem solving in mathematics, coding, vision, medicine, law, psychology and more. The study regards GPT-4 as an early general AI since it can achieve problem solving capabilities similar to humans in many cases.
Especially, ChatGPT has an excellent performance on coding tasks~\cite{dong2023self,tian2023chatgpt,biswas2023role} as it can generate code blocks in various programming languages based on either various formats of prompts such as incomplete code or description in natural languages. 
ChatGPT as a multi-purpose pretrained language model has a similar or superior performance on code generation compared with previous code-specific models.

As ChatGPT shows a remarkable performance on code generation, we naturally raise the question: What is the potential of ChatGPT for software security?
In the rapidly evolving landscape of software development, the urgency to maintain and enhance software security is paramount. Security vulnerabilities, software bugs, and inefficient patching mechanisms pose significant risks, making the detection, analysis, and rectification of these issues a critical area of study.
Recent studies conducted preliminary evaluations on the use of ChatGPT on security-related tasks.
There are experiments showing that ChatGPT may generate insecure code~\cite{liu2023your,khoury2023secure}, similar to previous code generation models~\cite{pearce2022asleep}.
A number of studies found that ChatGPT can also be used to resolve insecure code with proper prompts.
Pearce et. al.~\cite{pearce2022examining} used several LLMs earlier than ChatGPT and manually written prompts to repair security vulnerabilities. These LLMs resolved part of the test cases and the repair was unstable.
Sobania et. al.~\cite{sobania2023analysis} tested code repair on ChatGPT and drew a similar conclusion on ChatGPT.
Xia et. al.~\cite{xia2023conversational} emphasized the importance of feedback on LLM-based code repair. To be specific, after the LLM returned an invalid patch, they appended the result of test cases, which indicates why the fixed code failed, to the prompt and query the LLM for a second time. The work presented the possibility to use LLMs including ChatGPT to fix general program bugs.
On the other hand, security researchers try to leverage LLMs in security tools.
Deng et. al.~\cite{deng2022fuzzing} leveraged a LLM model to mutate the inputs (i.e., code evaluations by deep learning networks) when fuzzing deep learning libraries.
The above works illustrate several promising cases of using LLMs in security research or engineering but the aforementioned tasks do not cover broad software security and the latest ChatGPT is not widely tested.
To understand how much ChatGPT can help in the above security tasks, one should analyze ChatGPT's security-specific capabilities using representative test cases on a broad scope of software security tasks but such effort is in absence.

This work presents a comprehensive evaluation of the capabilities of ChatGPT in various software security tasks, as shown in \cref{fig:category_chart}, including vulnerability detection and repair, bug fixing, patching, software debloating, root cause analysis, decompilation, symbolic execution, and fuzzing. As these tasks traditionally require specialized knowledge and significant manual effort, the integration of AI-driven models like ChatGPT could greatly enhance efficiency and effectiveness in detecting and fixing software security.
For each software security task, we aim to answer a series of research questions:
\begin{itemize}
    \item Can ChatGPT understand the goal of the task?
    \item How accurate the ChatGPT's answers are?
    \item How different GPT-3.5 and GPT-4 performs?
    \item What is the limitation of ChatGPT if it cannot accomplish the task?
\end{itemize}

In terms of our methodology, we collected benchmark datasets for vulnerability detection, vulnerability repair, bug fixing and decompilation and used the datasets to systematically evaluate the performance of ChatGPT on such tasks, giving quantitative results.
For other software security tasks, we created representative test cases manually or from motivating examples of related works. These test cases are designed to demonstrate a certain aspect of ChatGPT's capability that is useful for security tasks.
For all experiments, we analyzed the outcomes of ChatGPT and discussed our hypothesis of ChatGPT's strengths and limitations.

In general, ChatGPT demonstrates impressive capabilities in software security tasks. With proper prompting, ChatGPT can easily understand the purpose of the tasks and generate reasonable responses.
Especially, ChatGPT with GPT-4 presents impressively high accuracy by solving 92\% of vulnerability detection cases, 95\% of vulnerability repair challenges, 84\% of bug fixing cases, and most manually crafted test cases across various tasks, which is a significant improvement compared with GPT-3.5.
Surprisingly, ChatGPT is even able to decompile the assembly language at least on short programs, illustrating that ChatGPT can process software context not limited to source code.
We also identify several limitations of ChatGPT. It cannot interpret binary or hexadecimal code and is limited to long code context. Such limitations compromise the usefulness of ChatGPT in real-world software development. For instance, the accuracy of vulnerability detection drops from 98\% to 66\% if ChatGPT is tested on real-world software vulnerabilities instead of synthetic programs. It is a promising research direction to address such limitations and systematically integrate ChatGPT into security-critical applications.

\mysection{Related work}

\textbf{Development of ChatGPT}.
Instruct-GPT~\cite{ouyang2022training} serves as the foundational large language model aligned with human feedback, upon which ChatGPT (GPT-3.5 and GPT-4) is built. As demonstrated in~\cite{ouyang2022training}, Instruct-GPT exhibits significant improvements over GPT-3 in NLP tasks such as question answering, reading comprehension, and summarization. OpenAI further enhanced the model by incorporating Reinforcement Learning from Human Feedback (RLHF)~\cite{christiano2017deep}, resulting in the more powerful and intelligent GPT-3.5-turbo (ChatGPT). Not only does GPT-3.5 excel in these tasks, but its outputs also align more closely with human responses.
When it comes to natural language tasks, ChatGPT (GPT-3.5) showcases evident improvements compared to previous LLMs. These advancements become even more apparent in GPT-4~\cite{gpt-4}. As illustrated in~\cite{gpt-4}, GPT-4 achieves remarkable success on diverse and challenging benchmarks, including more demanding academic and professional exams such as GRE, SAT, and AP exams. Surprisingly, GPT-4 demonstrates human-level performance on the majority of these exams, representing a substantial improvement over GPT-3.5. These results underscore the strong language understanding and reasoning abilities of ChatGPT (GPT-3.5 and GPT-4).

\textbf{Evaluation of ChatGPT's multi-task capability}.
Notably, ChatGPT (GPT-3.5)~\cite{chatgpt} and GPT-4~\cite{gpt-4} have opened up a highly promising avenue of research in artificial intelligence and machine learning. Numerous studies have investigated the capabilities of ChatGPT in different scenarios and tasks. Bubeck et. al.~\cite{bubeck2023sparks} demonstrates the impressive performance of GPT-4 in various categories of tasks, including coding, math, interaction with human, and so on.
The study argues that GPT-4 is already a general model that have capability comparable with human in diverse tasks.

For software security,
Several studies attempt to connect ChatGPT or other LLMs with software security and evaluate its ability in these tasks. 
Regarding vulnerability detection, Cheshkov et. al.~\cite{cheshkov2023evaluation} evaluated ChatGPT's ability to detect vulnerabilities in programs.
In the domain of secure code generation, Pearce et. al.~\cite{pearce2022asleep} evaluated whether Copilot~\cite{copilot} always generates secure code and concluded that Copilot generates insecure code with a high probability.
For vulnerability repair, Pearce et. al.~\cite{pearce2022examining} conducted pioneering research that assessed the capability of various LLMs on repairing vulnerable code examples. The results showed that LLMs may repair the vulnerabilities but the performance is unstable and is greatly affected by prompts, vulnerability types, programming languages, etc.
For automated program repair (APR) realm, 
Sobania et. al.~\cite{sobania2023analysis} and Xia et. al.~\cite{xia2023conversational} analyzed the ChatGPT's performance in bug fixing. By providing the buggy code and optional error messages, ChatGPT can resolve algorithmic bugs in some cases.

However, we find that existing evaluation studies cannot answer how useful ChatGPT is for software security tasks. The reason is twofold. 
First, existing studies mainly focus on the detection and repair of vulnerabilities or bugs, leaving the broad scope of software security untouched.
Second, many studies are accomplished before the release of the latest GPT-4 and thus cannot represent the state-of-the-art LLM performance.
To bridge the gap, we present a comprehensive evaluation study using the latest ChatGPT versions and considering broad software security tasks.  

\textbf{Applications built upon ChatGPT}.
A series of research aims to improve ChatGPT's capability in certain tasks by carefully designing the interaction with ChatGPT systematically.
MM-ReAct~\cite{yang2023mm} is a framework that integrated ChatGPT with a pool of vision experts to achieve visual reasoning and action. Though ChatGPT cannot directly process images and videos, MM-ReAct uses ChatGPT to intelligently choose what vision algorithm should be applied on which target.
Visual ChatGPT~\cite{wu2023visual} also integrated vision models and ChatGPT and provided an intelligent user interface.
GP-Tutor~\cite{chen2023gptutor} leveraged ChatGPT's strong code understanding capability to explain the functionality of source code for education purposes.
Park et. al.~\cite{park2023generative} used multiple ChatGPT instances to interact with each other and produced a world with generative agents where AI simulates the behavior of human.
There are also numerous other intriguing and powerful applications that are combined with ChatGPT including coding~\cite{dong2023self,feng2023investigating,liu2023improving,chen2023gptutor}, multimodality~\cite{bang2023multitask, yang2023mm, wu2023visual}, automatic multitask systems~\cite{park2023generative}, education~\cite{baidoo2023education}, and environmental applications~\cite{biswas2023potential}.

The integration of LLMs with traditional security tasks also presents a highly promising research direction.
Xia et. al.~\cite{xia2023automated} evaluated the performance of automated program repair (APR) using LLMs, demonstrating the considerable capability of LLMs with well-designed prompts.
Moreover, ChatRepair~\cite{xia2023keep} utilized conversation with ChatGPT to fix program bugs, i.e., providing ChatGPT with execution information as the feedback.
Jin et. al.~\cite{jin2023inferfix} proposed an end-to-end repair pipeline specifically tailored to address program bugs.
Deng et. al.~\cite{deng2023large} introduced FuzzGPT, which combines LLMs to automatically generate valid fuzzing input.
Hu et. al.~\cite{hu2023augmenting} proposed CHATFUZZ, a greybox fuzzer augmented by generative AI, demonstrating advancements in the field of fuzzing techniques.

Above ChatGPT-based applications indicate the potential of integrating ChatGPT in complicated and diverse tasks. In this work, we generalize ChatGPT to various software security tasks. Our design could be a baseline for future research of ChatGPT-based software security applications.                                 

\begin{table}[t]
\scriptsize
\vspace{0.10in}
\caption{Vulnerability detection on synthetic code.}
  \label{tab:cwe}
  \setlength{\tabcolsep}{4pt}
  \centering
  \begin{tabular}{| c | c | c | c | c | c | c | c | c | c |} \noalign{\global\arrayrulewidth1pt}\hline\noalign{\global\arrayrulewidth0.4pt}
        \multirow{2}{*}{ID}  &  \multirow{2}{*}{\#Case} & \multicolumn{4}{c|}{GPT-3.5} & \multicolumn{4}{c|}{GPT-4} \\
        \cline{3-10}
        &  & TP & FP & Precision & Recall 
        & TP & FP & Precision & Recall  \\
         \hline
    CWE-22 & 10 & 5 & 1 & 83.33\% & 100\% & 5 & 1 & 83.33\% & 100\%\\
    CWE-78 & 10 & 5 & 3 & 62.50\% & 100\% & 5 & 1 & 83.33\% & 100\%\\
    CWE-79 & 10 & 5 & 4 & 55.56\% & 100\% & 5 & 1 & 83.33\% & 100\%\\
    CWE-89 & 10 & 5 & 3 & 62.50\% & 100\% & 5 & 0 & 100\% & 100\%\\
    CWE-119 & 10 & 4 & 3 & 57.14\% & 80\% & 5 & 1 & 83.33\% & 100\%\\
    CWE-125 & 10 & 5 & 2 & 71.43\% & 100\% & 5 & 2 & 71.43\% & 100\%\\
    CWE-190 & 10 & 3 & 2 & 60\% & 60\% & 4 & 1 & 80\% & 80\%\\
    CWE-416 & 10 & 5 & 2 & 71.43\% & 100\% & 5 & 0 & 100\% & 100\%\\
    CWE-476 & 10 & 4 & 1 & 80\% & 80\% & 5 & 0 & 100\% & 100\%\\
    CWE-787 & 10 & 3 & 0 & 100\% & 60\% & 5 & 0 & 100\% & 100\%\\
    \hline
    \multicolumn{1}{|c|}{Total} & 100 & 44 & 21 & 67.69\% & 88\% & 49 & 7 & 87.50\% & 98\% \\
    \noalign{\global\arrayrulewidth1pt}\hline\noalign{\global\arrayrulewidth0.4pt}
  \end{tabular}
\end{table}

\begin{table}[t]
\scriptsize
\vspace{0.10in}
\caption{Vulnerability detection on CVEs.}
  \label{tab:cve}
  \setlength{\tabcolsep}{1.7pt}
  \centering
  \begin{tabular}{| c | c | c | c | c | c | c | c | c | c | c | c |}
    \noalign{\global\arrayrulewidth1pt}\hline\noalign{\global\arrayrulewidth0.4pt}
    
    \multirow{2}{*}{Language}  &  \multirow{2}{*}{\#Case} & \multicolumn{5}{c|}{GPT-3.5} & \multicolumn{5}{c|}{GPT-4} \\
    \cline{3-12}
    &  & TP & FP & Fail & Precision & Recall 
    & TP & FP & Fail & Precision & Recall  \\
     \hline
   
    C & 34 & 3 & 1 & 6 & 75\% & 21.43\% & 9 & 3 & 0 & 75\% & 52.94\%\\
    Cpp & 6 & 0 & 2 & 2 & 0\% & 0\% & 1 & 2 & 0 &  33.33\% & 33.33\%\\
    Python & 10 & 0 & 2 & 0 & 0\% & 0\% & 1 & 1 & 0 & 50\%& 20\% \\
    Go & 8 & 0 & 0 & 0 & 0\% & 0\% & 3 & 0 & 0 & 100\% & 75\% \\
    JavaScript & 8 & 2 & 2 & 0 & 50\% & 50\% & 2 & 0 & 0 & 100\% & 50\% \\
    PHP & 2 & 0 & 0 & 0 & 0\% & 0\% & 1 & 0 & 0 &  100\% & 100\%\\
    \hline
    \multicolumn{1}{|c|}{Total} & 68 & 5 & 7 & 8 & 41.67\% & 17.24\% & 17 & 6 & 0 & 73.91\% & 50\% \\
    \noalign{\global\arrayrulewidth1pt}\hline\noalign{\global\arrayrulewidth0.4pt}
  \end{tabular}
\vspace{-0.10in}
\end{table}

\mysection{Vulnerability Detection}\label{sec:vul_detect}

Vulnerability detection is a task identifying and localizing the vulnerabilities which can be exploited by attacks in the software project systems.
Various detection methods have been proposed in the rich literature.
Conventional vulnerability detection methods~\cite{ko1994automated,xiao2020mvp,jeong2019razzer,codeql,infer,clang,hu2021automated} is usually based on static program analysis for matching predefined vulnerability patterns or dynamic methods which reveal vulnerabilities from the runtime execution traces. These methods are successful but may have limited generality, e.g., some solutions only work on certain programming languages or vulnerabilities. They may also suffer from inaccuracies as the design often depends on manual analysis of the vulnerabilities.
Machine learning based methods~\cite{li2020automated,russell2018automated,zagane2020deep,li2021automated,mirskyvulchecker} formulate the vulnerability detection as a classification problem. The models make binary decisions (vulnerable or not) by accepting code in plain text or certain structured representation, according to the learning on a large corpus. However, deep learning based solutions can also be limited in accuracy because the training dataset could be imbalanced or biased.

As large models like ChatGPT push language processing to a new level, it is promising to leverage ChatGPT to identify vulnerabilities by directly examining the source code.
Though vulnerability detection can also be applied to binary code or other intermediate representations (e.g., LLVM), we focus on source code because (1) the source code is in a form close to natural language which is originally used to train LLMs thus ChatGPT is supposed to perform better on source code instead of binary code; (2) the source code is always available during the software development thus it is easier to access than binary code.
Therefore, if ChatGPT performs well on source-code-level vulnerability detection, it is unnecessary to use ChatGPT to process binary code.

\textbf{Datasets}. We collect two datasets for evaluation. One is the dataset of synthetic code samples, compiled by 100 selected test cases from SARD dataset~\cite{209211}. The test cases are associated with 10 of the most popular vulnerabilities documented by Common Weakness Enumeration (CWE)~\cite{cwe}, including the notorious SQL injection (CWE-89), cross-site scripting (CWE-79), path traversal (CWE-022), etc. 10 test cases are prepared for each CWE, including 5 vulnerable cases and 5 benign cases. The test cases cover various languages including C, C++, Java, and Python.

The other dataset is about real-world vulnerabilities, which is compiled from exploitable vulnerabilities in open-sourced software. We select 34 vulnerabilities from the database of Common Vulnerabilities and Exposures (CVE), each with the vulnerable version of the software and the official fixed version of the software. Our collection covers 9 vulnerabilities defined by CWE and 6 different programming languages.

\textbf{Prompts}. To ask ChatGPT for detecting vulnerabilities, we use a prompt as ``Does the following code have vulnerabilities? \{code\}'' where ``\{code\}'' is the placeholder of the source code. 
For real-world vulnerabilities, as ChatGPT can accept the limited length of the input, we cannot directly feed the whole software to ChatGPT. Instead, we select vulnerable functions in the software as the input.

\textbf{Metrics}. We manually confirm whether the vulnerability detection is successful. We extract the names of vulnerabilities (if any) from the response of ChatGPT and regard the detection as successful when (1) the detected vulnerabilities include the ground-truth vulnerability in vulnerable cases, or (2) no vulnerability is detected in benign cases. We first label each test case as false positive (FP), false positive (FP), true negative (TN), or false negative (FN), according to the ground truth and the detection result.
Then we calculate precision ($\frac{TP}{TP + FP}$) and recall ($\frac{TP}{TP + FN}$) which are widely used in previous studies~\cite{xiao2020mvp,olson2008advanced,powers2020evaluation}. Higher precision and higher recall indicate more accurate detection.

In the following subsections, we will present and analyze ChatGPT's performance of vulnerability detection on the synthetic dataset and the real-world dataset, respectively.

\textbf{Detecting vulnerabilities in synthetic programs}.
The main results are shown in~\cref{tab:cwe}. ChatGPT is impressive in detecting vulnerabilities in synthetic code cases. GPT-3.5 successfully detects 44 out of 50 vulnerable cases, which is a precision of 67.69\%. It failed on 6 vulnerable test cases, having a recall rate of 88\%. 
GPT-4 detects 49 out of 50 vulnerable cases in a precision of 87.5\% and has only one false negative case.
GPT-4 greatly outperforms GPT-3.5 by 19.81\% in precision and it reaches 98\% in recall rate. The result demonstrates the strong capability of ChatGPT, especially with GPT-4, on detecting vulnerabilities. In terms of the inaccuracies, ChatGPT suffers from false positives in vulnerability detection as the precision is only 67.69\%/87.50\% for GPT-3.5/GPT-4. The problem of false negatives is relatively minor, at least on short synthetic programs, as the recall reaches 88\%/98\% for GPT-3.5/GPT-4. We use case studies to analyze both successful cases and inaccurate detection results as follows.

\begin{figure}[t]
    \centering
    \includegraphics[width=0.49\textwidth]{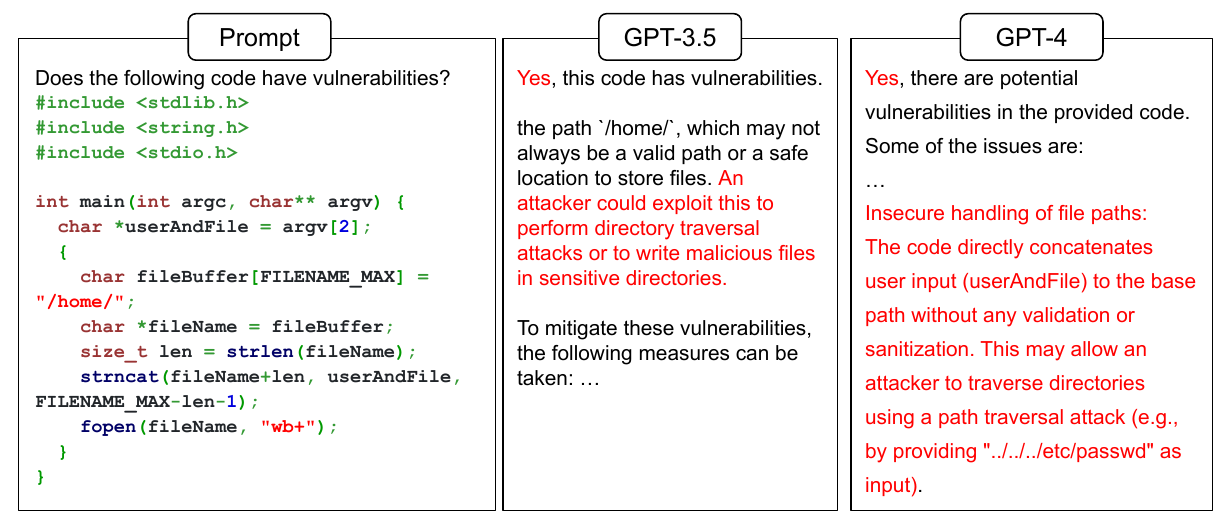}
    \caption{One successful example of ChatGPT's vulnerability detection (path traversal in C program).}
    \label{fig:detect1}
\end{figure}


\begin{figure}[t]
    \centering
    \includegraphics[width=0.49\textwidth]{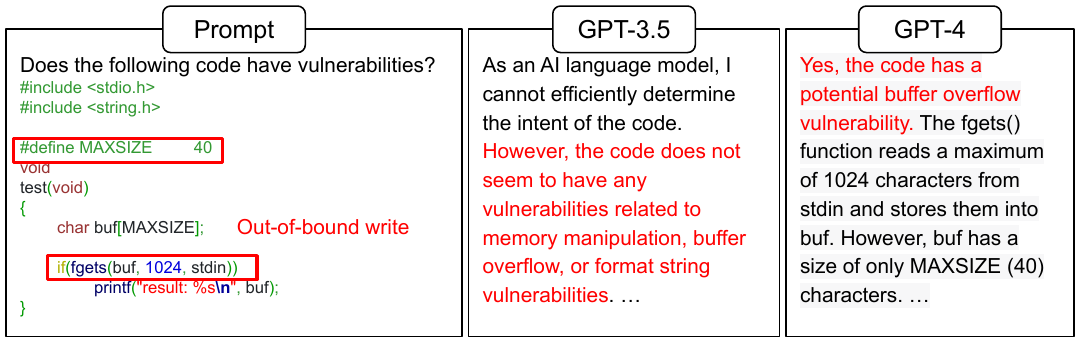}
    \caption{One false negative example of ChatGPT's vulnerability detection (Out-of-bounds Write in C program).}
    \label{fig:detect6}
\end{figure}




We first study one successful example to demonstrate how ChatGPT resolves vulnerability detection challenges. As shown in~\cref{fig:detect1}, the example involves the path traversal vulnerability. The code opens a file whose path is determined by user-provided values thus the attacker may illegitimately access unexpected files using malicious inputs. ChatGPT first confirm the existence of vulnerabilities by saying ``yes''. Then it explains the mechanism of the path traversal and precisely identifies the key vulnerable variable. In addition, it recommends mitigation methods for the vulnerabilities.
ChatGPT is also able to discover multiple vulnerabilities in one pass, as shown in GPT-4's result in~\cref{fig:detect1}. Some discovered vulnerabilities are minor issues but are associated with certain inappropriate coding styles. These findings also benefit the robustness of the code.
As demonstrated by the example, ChatGPT's comprehensive responses can greatly help developers to build robust code. It not only warns developers with security vulnerabilities but also proposes actionable solutions. 

One major cause of inaccurate detection is ChatGPT's limited precision on understanding numerical operations.
 As an example of false negatives, the code in \cref{fig:detect6} contains a buffer overflow vulnerability because at most 1024-byte input from \emph{stdin} is stored into an array in only a 30-byte length. However, ChatGPT does not detect the fact that 30 bytes is much smaller than 1024 bytes.
Because of this limitation, ChatGPT has relatively worse performance on overflow vulnerabilities (CWE-119, CWE-125, and CWE-190) among all tested vulnerabilities, as shown in~\cref{tab:cwe}.

\textbf{Detecting vulnerabilities in real software projects}.
Ideally, vulnerability detection is going to be involved in the development of real complex software projects, which is more challenging as these projects often involves context non-trivial to understand and complicated dependencies of libraries or modules.
It is critical to study whether ChatGPT can end-to-end address real-world vulnerabilities.

The results are shown in the~\cref{tab:cve}. Precision and recall drop significantly compared with the results on synthetic programs (\cref{tab:cwe}), which indicates the additional challenges on detecting vulnerabilities on complex software instead of short programs. Especially, ChatGPT with GPT-3.5 is almost unusable as it produces only 41.67\% precision and 17.42\% recall. Results from GPT-4 is better (73.91\% precision and 50\% recall) but it still suffers from a high number of false positives and false negatives. We analyze the failing cases as follows.

\begin{figure}[t]
    \centering
    \includegraphics[width=0.49\textwidth]{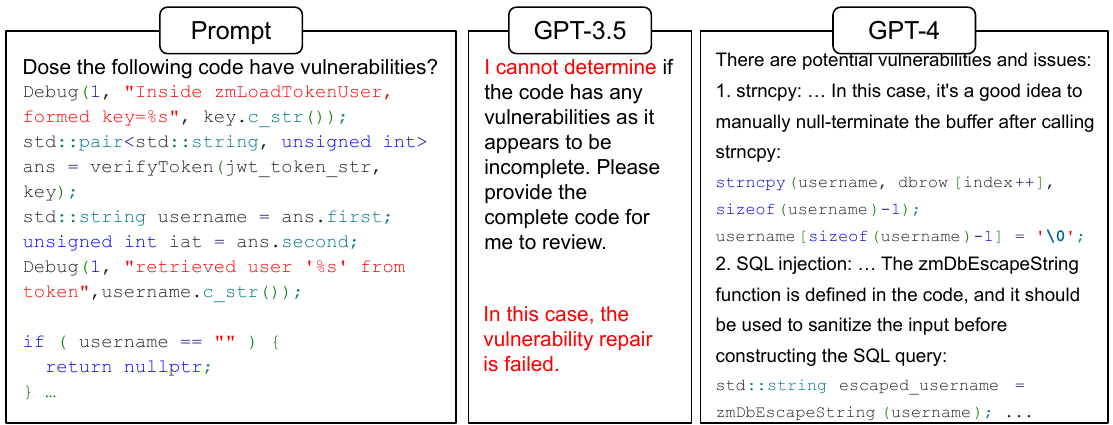}
    \caption{One failure in ChatGPT's vulnerability detection on real-world vulnerability (CVE-2023-26032).}
    \label{fig:detect2}
\end{figure}

\begin{figure}[t]
    \centering
    \includegraphics[width=0.49\textwidth]{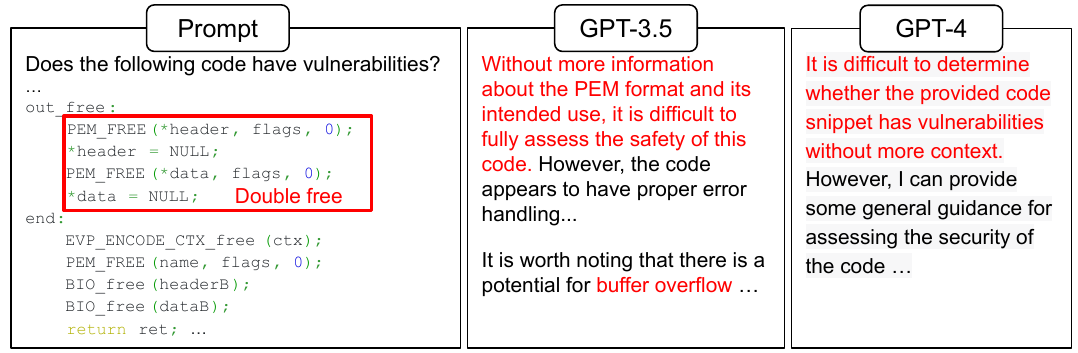}
    \caption{One false negative of ChatGPT's vulnerability detection on real-world vulnerability (CVE-2022-4450).}
    \label{fig:detect7}
\end{figure}

\emph{Nondeterministic results}.
ChatGPT may produce nondeterministic results saying that the code is over-complicated or incomplete to analyze, labeled as ``failed'' in \cref{tab:cve}. As shown in~\cref{fig:detect2}, when we ask ChatGPT with GPT-3.5 to examine the vulnerable function (a SQL injection vulnerability), it complains that the code is incomplete, which is likely caused by undefined identifiers and undeclared dependencies. However, it is extremely hard to extract a standalone piece of code from a real project with complex dependencies. On contrast, ChatGPT with GPT-4 still succeed in detecting the vulnerability in this case. 
Though the complicated code dependencies are causing failures, the success of GPT-4 indicates the potential of ChatGPT to process incomplete code.

\emph{False negatives}.
Compared with vulnerability detection on short synthetic programs, the detection on real software projects suffers more significantly from false negatives. We hypothesize that it is because the complex code logic obscures the pattern of the vulnerabilities. For example, as shown in~\cref{fig:detect7}, CVE-2022-4450 contains a double free vulnerability (CWE-415) where it frees the variable \emph{header} and \emph{data} twice. Though the pattern of the double free could be as simple as a simple control dependency, the two free operations could be located in two far-away locations in the program and hidden among a large body of code. Because of complex context, ChatGPT fails to identify the exact vulnerability and instead offers some minor suggestions on general sanity checks.

\mysection{Vulnerability Repair}
\label{sec:vul_repair}


To address security vulnerabilities, one major solution is manual patching by the developers. The developers of the vulnerable software analyze the root cause of the vulnerabilities and provide a modification on the source code (i.e., patch) to fix the vulnerabilities. 
Then the developers will release the fixed version and 
The users of the software can then apply the patch or update the software to a safe version. However, this process requires significant expert knowledge and is time-consuming.
Automatic vulnerability repair~\cite{xu2020automatic,christou2022ivysyn,mulliner2013patchdroid,arnold2009ksplice,duan2019automating} is proposed to achieve timely fix of vulnerability with a minimum of human effort. 
Conventional methods of vulnerability repair rely on manually defined rules or evolutionary search of patches, which are limited in flexibility and scale.

As vulnerability repair aims to generate safe code given the vulnerable code, it is basically a language generation task based on prior knowledge, which is a suitable use case for LLMs like ChatGPT. 
Recent studies analyze vulnerability repair using LLMs and provide preliminary results.
Pearce et al.~\cite{pearce2022examining} manually crafts prompts to perform vulnerability repair and tested multiple LLMs (but not including ChatGPT). Their analysis revealed that LLM-based code repair can deliver comparable performance to conventional methods, but its stability is limited by various factors, such as the nature of the vulnerabilities and the quality of the prompts used.
Based on the existing studies, we analyze the capability of vulnerability repair with ChatGPT to provide a comprehensive analysis of its strengths and limitations.

\textbf{Datasets}. We set two different test datasets, following the same methodology in~\cref{sec:vul_detect}.
The test cases include the synthetic test cases from SARD and Juliet~\cite{black2018juliet} and 2) the real CVE vulnerabilities in open-source software.
For the synthetic dataset, we choose 18 CWEs from MITRE's ``2021 CWE Top 25'' list following~\cite{pearce2022asleep} and compile 154 test cases in total. 
For real-world vulnerabilities, we collect recent CVEs (starting from September 2019) on OpenSSL~\cite{openssl} as the test data.  

\textbf{Prompts}. We use the most basic prompt, a straightforward question ``Can you fix the vulnerabilities in the following code?''. 


\textbf{Metrics}. We manually validate repairs generated by ChatGPT. One repair is successful if the vulnerability is no longer exploitable. We calculate the success rate of vulnerability repair as the evaluation metric.

\begin{table*}[!htb]
\minipage{0.31\textwidth}
\small
\vspace{0.10in}
\caption{Vulnerability fix on CWEs.}
    \tiny
  \label{tab:fix_cwe}
  \centering
  \begin{tabular}{| c | c | c | c | c |}
    \noalign{\global\arrayrulewidth1pt}\hline\noalign{\global\arrayrulewidth0.4pt}
     {CWE ID} & {GPT-4} & {GPT-3.5} \\
    \hline
    CWE-20 & 3/3  & 3/3   \\
    CWE-22 & 13/13  & 8/13   \\
    CWE-78 & 13/13  & 11/13   \\
    CWE-79 & 11/13  & 10/13   \\
    CWE-89 & 13/13  & 11/13  \\
    CWE-119 & 13/13  & 13/13   \\
    CWE-125 & 12/13  & 10/13   \\
    CWE-190 & 10/13  & 10/13   \\
    CWE-200 & 0/3  & 0/3  \\
    CWE-306 & 2/3  & 0/3   \\
    CWE-416 & 12/13  & 11/13  \\
    CWE-434 & 2/3  & 3/3   \\
    CWE-476 & 12/13  & 13/13 \\
    CWE-502 & 3/3  & 3/3 \\
    CWE-522 & 3/3  & 1/3  \\
    CWE-732 & 3/3 & 2/3   \\
    CWE-787 & 12/13  & 9/13  \\
    CWE-798 & 2/3  & 1/3  \\
    \hline
    {Total} & 139/154  & 119/154  \\
    \noalign{\global\arrayrulewidth1pt}\hline\noalign{\global\arrayrulewidth0.4pt}
  \end{tabular}
\endminipage
\minipage{0.66\textwidth}
\small
\vspace{0.10in}
\caption{Vulnerability fix on CVEs.}
\scriptsize
  \label{tab:fix_cve}
  \centering
  \begin{tabular}{| c | l | c | c | c | c |}
    \noalign{\global\arrayrulewidth1pt}\hline\noalign{\global\arrayrulewidth0.4pt}
    {ID} &{Vulnerability} & {GPT-4} & {GPT-3.5} \\
 
    \hline
    CVE-2023-0216 & NULL dereference triggered by malformed PKCS7. & $\surd$  & $\surd$  \\
    CVE-2023-0401 & NULL dereference triggered by wrongly digested PKCS7. & $\times$  & $\surd$ \\
    CVE-2023-0217 & NULL dereference in BN\_copy. & $\surd$  & $\surd$  \\
    CVE-2022-4450 & Double free on *header and *data in PEM\_read\_bio\_ex(). & $\surd$ & $\times$  \\
    CVE-2022-3996 & Redundant lock operation in x509. & $\times$  & $\times$  \\
    CVE-2022-3602 & Potential buffer overflow in ossl\_a2ulabel(). & $\times$  & $\times$  \\
    CVE-2022-3358 & NULL dereference in EVP\_CIPHER. & $\times$ & $\times$  \\
    CVE-2022-29242 & Potential buffer overflow in GOST key exchange. & $\times$ & $\times$  \\
    CVE-2022-1434 & Wrong crypto in RC4-MD5. & $\times$ & $\times$ \\
    CVE-2022-1343 & Improper OCSP\_basic\_verify signer certificate validation. & $\times$ & $\times$ \\
    CVE-2022-0778 & Possible infinite loop in BN\_mod\_sqrt(). & $\times$ & $\times$\\
    CVE-2021-4044 & Possible infinite loop in X509\_verify(). & $\times$ & $\times$  \\
    \hline
    \multicolumn{2}{|c|}{Total} & 3/12 & 3/12  \\
    \noalign{\global\arrayrulewidth1pt}\hline\noalign{\global\arrayrulewidth0.4pt}
  \end{tabular}
\endminipage\hfill
\vspace{-0.20in}
\end{table*}

\textbf{Repairing vulnerabilities in synthetic programs}.
We 
introduce the overall results presented in~\cref{tab:fix_cwe}. 
GPT-3.5 successfully repairs 119 out of 154 (77.27\%) test cases without any hints about the vulnerabilities. 
As for GPT-4, which can repair 139 out of 154 test cases without vulnerability information.
GPT-4 showcases a much stronger capability for vulnerability repair than GPT-3.5.

\begin{figure}[t]
    \centering
    \includegraphics[width=0.49\textwidth]{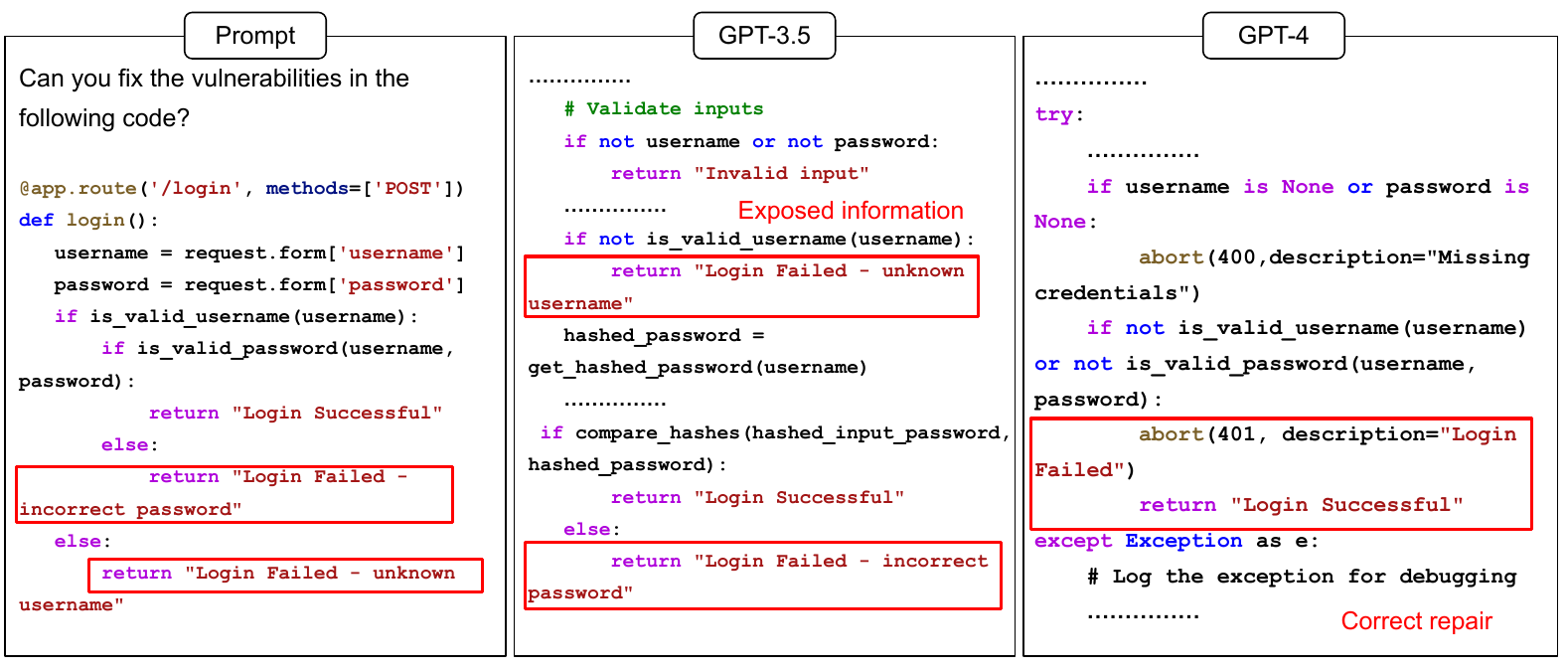}
    \caption{One failed case of ChatGPT's vulnerability repair on exposure of sensitive information (CWE-200).}
    \label{fig:cwe-200}
\end{figure}

We use a case study to demonstrate the different performance of GPT-3.5 and GPT-4.
As shown in \cref{fig:cwe-200}, we want to repair a login service code that contains an exposure of sensitive information.
The error messages sent to users reveal sensitive information, including the reason of the login failures (e.g., ``unknown username'' and ``incorrect password''), which could be leveraged by attackers to exploit the system. 
While GPT-3.5 fails to address this vulnerability effectively, GPT-4 demonstrates its success in mitigating the issue and safeguarding sensitive information.

\textbf{Repairing vulnerabilities in real software projects}.
As shown in the table, ChatGPT can fix most NULL dereference vulnerabilities (e.g., CVE-2023-0216 and CVE-2023-0217).
It can hardly address other vulnerabilities, especially the ones involving complex application contexts. For instance, the buffer overflow of CVE-2022-29242 is hidden in a complex code with more than 3000 lines and involves the logic of key exchange protocol, which is hard for ChatGPT to process.
As a result, ChatGPT repairs 3 and 4 out of 12 cases with GPT-3.5 or GPT-4 respectively.
Extracting informative code snippets and messages from complex code as part of the prompt poses challenges, but it holds great promise for future advancements.

\mysection{Bug Fixing}
\label{sec:bug_fixing}

\begin{figure}[t]
    \centering
    \includegraphics[width=0.49\textwidth]{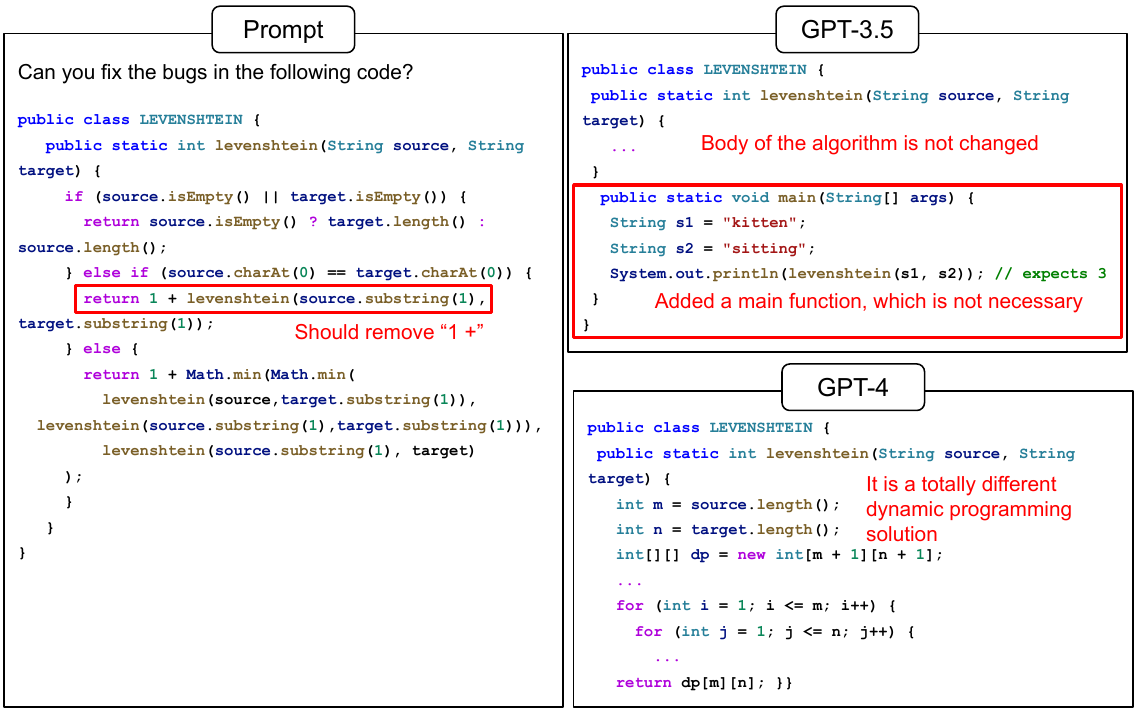}
    \caption{One failed example of ChatGPT's bug fixing on an implementation of levenshtein algorithm.}
    \label{fig:bug1}
\end{figure}


Software bugs can cause errors, crashes, and other issues that can affect the functionality, performance, and security of software systems.
Automatic bug fixing is the process of automatically identifying and repairing software bugs without human intervention.
Different from vulnerability repair where vulnerabilities are predefined code patterns or behaviors, bug fixing is usually provided with test cases that are specific for the program to test. A program is regarded as buggy if any of the test cases fails. Automatic bug fixing aims to modify the source code to remove the bug.

Automatically fixing bugs is a hot topic in software engineering, which is also referred as Automated Program Repair (APR). Traditional APR~\cite{ghanbari2019practical,hua2018sketchfix,liu2019tbar,martinez2016astor, long2015staged, demarco2014automatic} is based on program synthesis which mutates the code until a good solution is found.
Machine learning based methods are also introduced. For instance, Neural machine translation (NMT)~\cite{chen2019sequencer,li2020dlfix,ye2022neural,zhu2021syntax} is a paradigm that could be trained to understand code structures and models based on NMT could be granted with code repair capabilities. 
As LLMs like ChatGPT demonstrate the strong capability of code understanding, recent studies proposed LLM-based bug fixing~\cite{sobania2023analysis, xia2023conversational}. They provide ChatGPT with the source code as well as manual feedback or results of test cases when the repair fails. Their evaluation illustrates that ChatGPT for bug fixing not only is possible but also outperforms various existing methods. However, existing discussion on ChatGPT's bug fixing is either on small scale or does not include the latest GPT-4.
Therefore, we will investigate the capability of ChatGPT with the latest GPT-4 version on a large benchmark.

\textbf{Datasets}.
We use QuixBugs\cite{lin2017quixbugs}, one of the most commonly used dataset of buggy programs. It contains 40 buggy Python programs and 40 buggy Java programs. Each buggy program has a set of test cases which indicates the expected behavior of the program.

\textbf{Prompts}.
We construct the prompt as ``Can you fix the bugs in the following code? \{code\}'' where ``\{code\}'' is the placeholder for the source code.

\textbf{Metrics}. We run the test cases given by QuixBugs to evaluate bug fixing results. A bug is successfully fixed if all test cases pass.



\textbf{Results}. GPT-3.5 fixes 24 out of 40 Python bugs and 14 out of 40 Java bugs. GPT-4 fixes 33 out of 40 Python bugs and 34 out of 40 Java bugs.
In general, ChatGPT could be used for bug fixing. 
Regarding GPT-3.5, its ability to repair code is limited and varies depending on the programming language. For instance, GPT-3,5 can fix 60\% Python bugs but only 35\% Java bugs.
GPT-4 in general yields better repair results. It achieves a success rate of 85\% and 82.5\% when applied to Java and Python test code, respectively. This underscores GPT-4's superior ability to comprehend and analyze buggy algorithmic code.

As an example of failed repairs, the code shown in~\cref{fig:bug1} is the Levenshtein algorithm written in Java. Levenshtein distance is the minimum number of single-character edits (insertions, deletions, or substitutions) required to change a string into the other one. The buggy code is the recursion form of the algorithm, when two char in the compared string equal, it adds an additional 1 into the total count which is a logic flaw (line 12).
Upon the request of bug fixing, GPT-3.5 adds an additional main function. Though the repair indeed makes the code more complete, it is not relevant to the bug.
GPT-4 addresses the bug but it changes the whole algorithm pattern. It generates a totally different implementation of the Levenshtein algorithm using Dynamic Programming (DP) instead of the original version using recursion.
In general, repairs from ChatGPT may contain various problems, including unnecessary changes, modifying original code logic, introducing new errors, etc.

\mysection{Patching}
\label{sec:patching}

Patching is a technique used in software engineering~\cite{monperrus2018automatic}, which involves applying a small piece of code, known as a patch, to an existing software system to correct an error or improve its functionality.
For software security especially, patching aims to resolve security vulnerabilities in the software system.

In a common situation, when a new vulnerability is revealed, the developers of the affected software obtain the vulnerability report and offer an official patch for the software. Other developers who used the software can shut down the current service, download and apply the official patch, then recompile and restart the system. However, in the real production environment, applying the official patches sometimes is not an available option in some situations. 
For instance, if the software is running and is too critical to shut down, developers need a hot patch that addresses the vulnerabilities while keeping the system running.
In addition, many services use old versions of the software which may not be maintained anymore, thus the official patch may conflict with the old version. Upgrading the software may be impractical thus the developers need to prepare their custom patch.

Given the above challenges of patching, recent studies~\cite{shi2022backporting, xu2020automatic,sidiroglou2005countering,martinez2015mining} propose to automatically transform the official patch to an applicable patch. Conventionally, the transformation is based on program analysis that suffers from high deployment complexity. 
Additionally, learning-based methods~\cite{kim2013automatic, ma2017vurle} leverage neural network models to generate secure patches similar to human-written patches.
With the development of LLMs,
ChatGPT is a possible alternative to end-to-end patching. We demonstrate ChatGPT-based patching in two real-world scenarios.
Note that we label deleted lines in red and inserted lines in our figures.

\begin{figure}[t]
    \centering
    \includegraphics[width=0.49\textwidth]{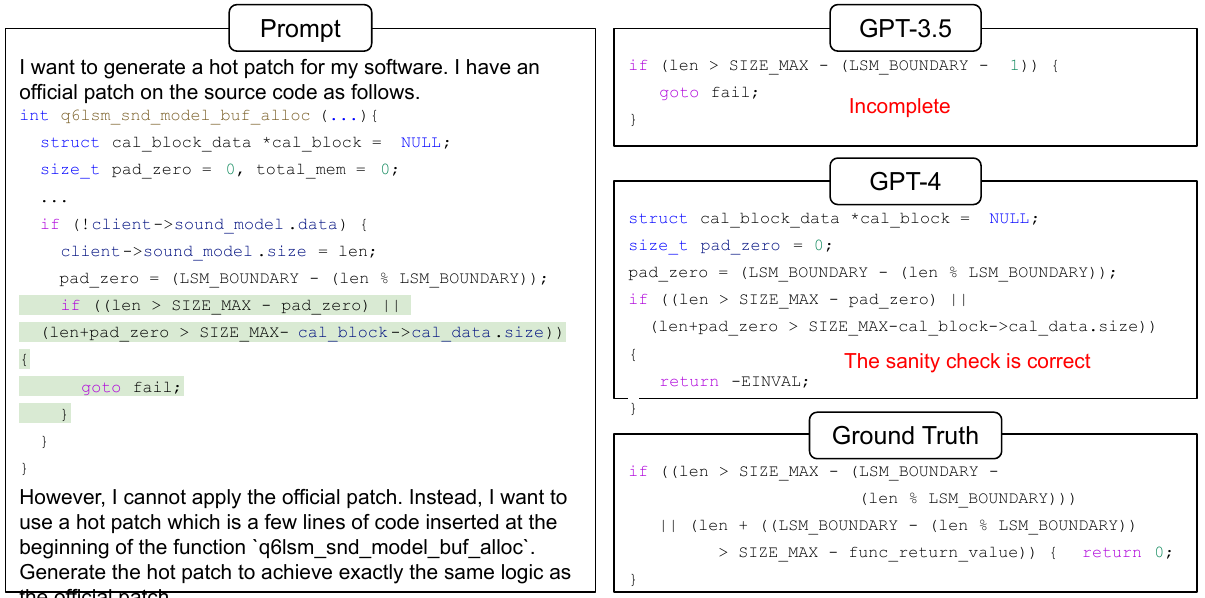}
    \caption{Ask ChatGPT to generate a hot patch for CVE-2015-8940 on Android kernel according to the official patch.}
    \label{fig:patching_1}
\end{figure}

\textbf{Hot patch generation}.
Hot patches aim to fix security vulnerabilities without stopping the service. Developers can hook the vulnerable function and apply a pre-constructed hot patch to it. In this way, the code defined by the hot patch is run before reaching the vulnerability. We inherit the definition of hot patch generation from Xu et. al.~\cite{xu2020automatic}:

\emph{Given a vulnerable function $F$ and its official patch $P$ at location $L$, we would like to find a suitable location $L'$ of $F$ in binary form to insert an automatically generated hot patch $P'$, which has the same semantics as $P$.}

We also use the motivating example in \cite{xu2020automatic} to demonstrate whether ChatGPT can handle the task of hot patch generation. For CVE-2015-8940 in Android Qualcomm msm kernel 3.10, there is an official patch as shown in ~\cref{fig:patching_1} which adds a sanity check (labeled in green) in function \emph{q6lsm\_snd\_model\_buf\_alloc}.
The hot patch should implement exactly the same logic as the sanity check but can only use variables available at the beginning of the function, i.e., function arguments (\emph{client} and \emph{len}), global function calls (\emph{cal\_utils\_get\_only\_cal\_block}), and global constants (\emph{SIZE\_MAX} and \emph{LSM\_BOUNDARY}).
We explain the concept of hot patch generation in English and use it as the prompt for ChatGPT.

ChatGPT can understand the request and produce reasonable hot patches but the generated ones may have problems. ChatGPT with GPT-3.5 misses part of the condition in the sanity check, resulting in an incomplete fix. ChatGPT with GPT-4 produces a good sanity check but also involves redundant check on \emph{cal\_block == NULL}, which exists in the original code. From the results, ChatGPT can generate reasonable and fairly good hot patches according to the user's requests, but the outcomes certainly need validation by either manual effort or automatic analysis. 

\begin{figure}[t]
    \centering
    \includegraphics[width=0.49\textwidth]{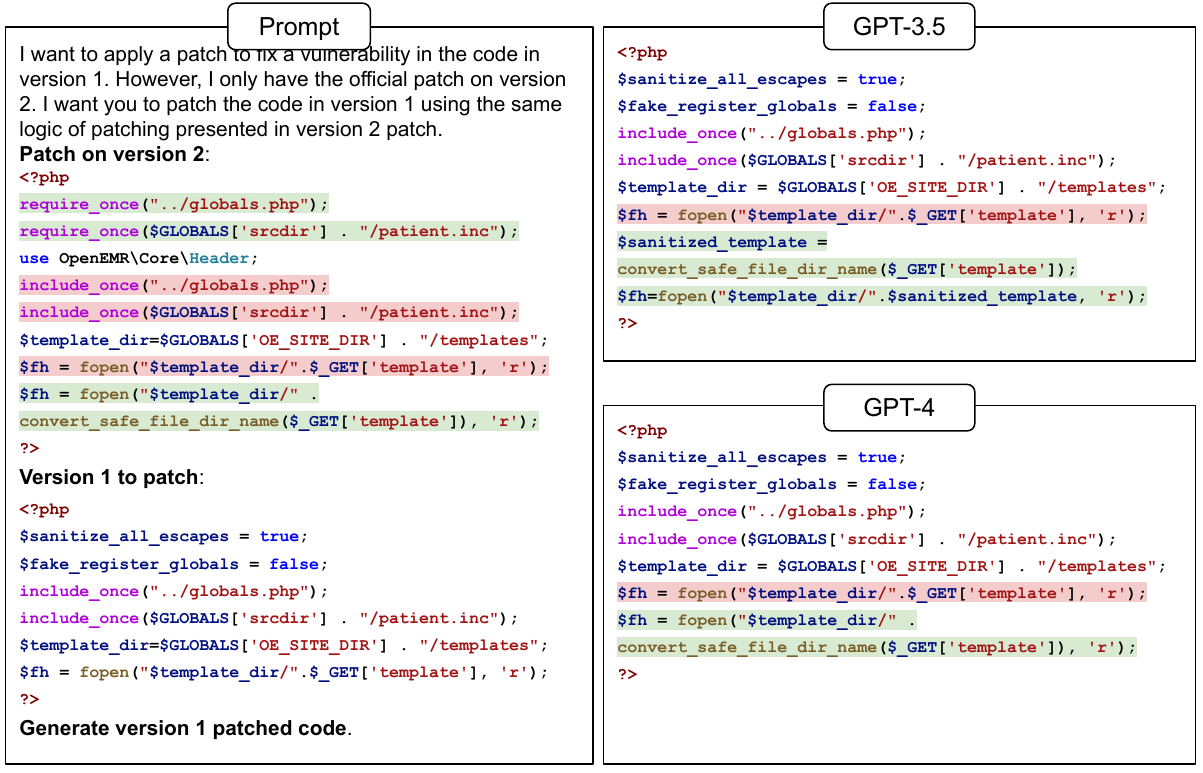}
    \caption{Ask ChatGPT to patch CVE-2015-8940 on OpenEMR 5.0.0.5 based on the official patch on 5.0.0.6.}
    \label{fig:patching_2}
\end{figure}

\textbf{Patch backporting}.
Patch backporting is a task for applying official patches in a newer version to software in an older version, as defined in \cite{shi2022backporting}. In the end, one should modify the official patches to avoid any conflict with the older-version software. For instance, if the official patch modifies a feature that is newly introduced after the older version, we should remove such modification from the patch. We use the motivating example in \cite{shi2022backporting} to demonstrate ChatGPT's capability in this task.

The example is about CVE-2018-10572 on OpenEMR, a software for managing medical records. The official patch as shown in ~\cref{fig:patching_2} is applied on OpenEMR 5.0.0.6 while we are going to patch OpenEMR 5.0.0.5. The key fix implemented by the official patch is the check on user-provided directory variable \emph{template}, in order to avoid path traversal vulnerabilities. The official patch on OpenEMR 5.0.0.6 also involves a few importing operations that are not relevant to the 5.0.0.5 version. Therefore, the patch for older version 5.0.0.5 should include the directory check but ignores the check on import statements. 

ChatGPT does a good job on this task. ChatGPT with either GPT-3.5 or GPT-4 generates a solid patch satisfying the aforementioned requirements. ChatGPT demonstrates a strong ability of understanding the context of code.                                
\mysection{Root Cause Analysis}
\label{sec:root_cause_analysis}

Root cause analysis is a technique used in software security to identify the underlying causes of security incidents, vulnerabilities, or breaches. The goal of root cause analysis is to determine why a security issue occurred and to develop solutions that address the underlying causes, rather than just treating the symptoms.

Existing tools~\cite{yagemann2021arcus,musuvathi2008finding,ahmed2014impact,blazytko2020aurora,attariyan2012x,cui2018rept,cui2016retracer,xu2017postmortem} for root cause analysis first gather information about the errors or vulnerabilities, such as bug messages from the operating system or compilers, execution traces from debuggers, etc.
Then program analysis methods are applied to analyze collected data and identify root causes, which could be in various graininess such as specific function, a code block, or one improper sanity check.

Basically, root cause analysis is a task of understanding the causal relation between the code and the error or vulnerability. Such context learning could be a strength of LLMs like ChatGPT. In this section, we analyze ChatGPT's capability in root cause analysis using case studies. We consider using either error messages or test cases for identifying the root cause.

\begin{figure}[t]
    \centering
    \includegraphics[width=0.49\textwidth]{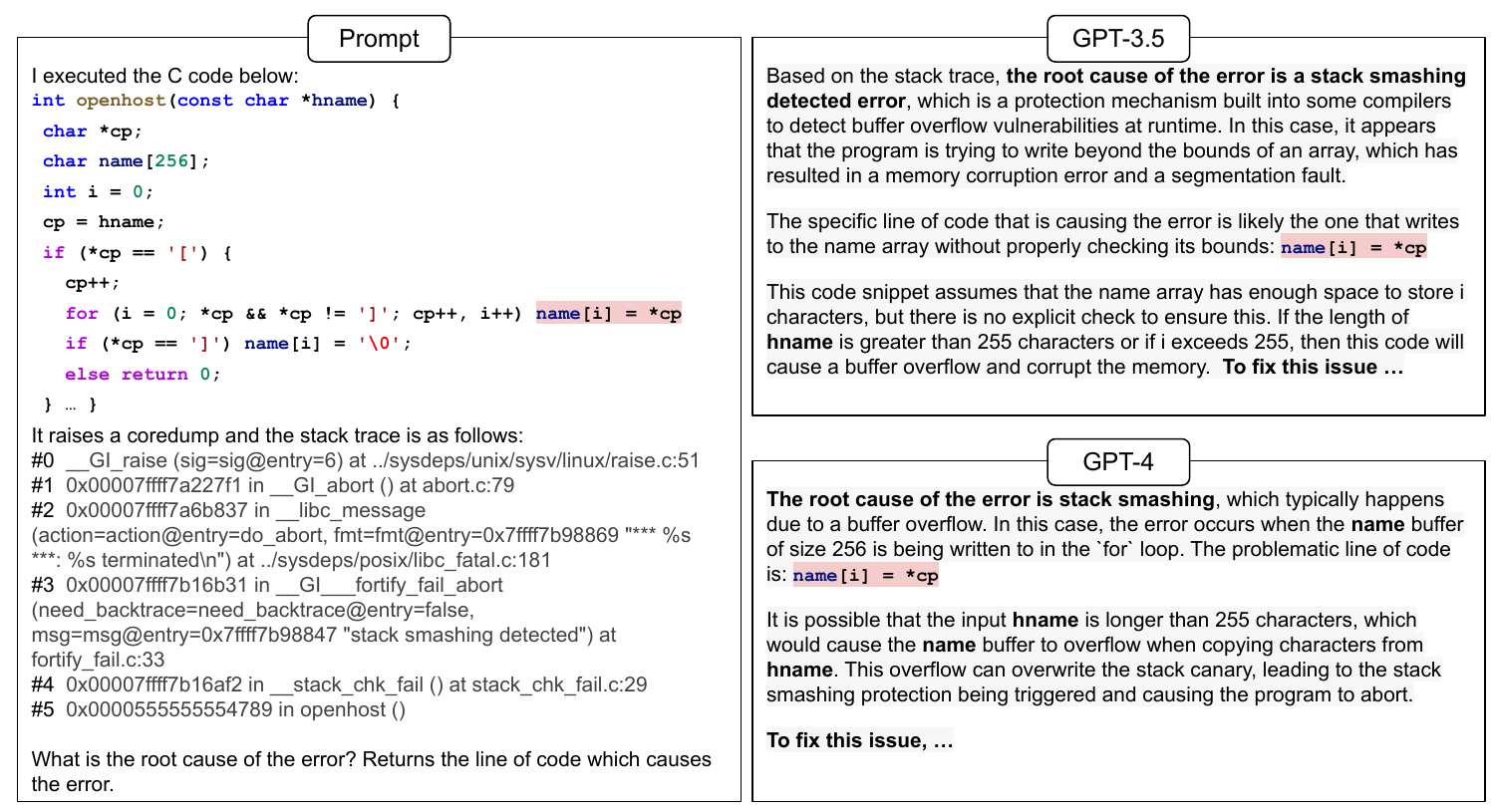}
    \caption{Given the error message (stack trace), find the root cause of stack overflow in \emph{ntpq} (CVE-2018-12327).}
    \label{fig:root_cause_analysis_1}
\end{figure}

\textbf{Root cause analysis based on error messages}.
As a case study, we use the vulnerability of CVE-2018-12327 in \emph{ntpq}, which is also the motivating example in \cite{yagemann2021arcus}. CVE-2018-12327 is a stack overflow vulnerability in \emph{ntpq} to achieve arbitrary code execution because it lacks a sanity check for the length of the relevant command line argument.

In the first case study, we assume that an exploit is found during the execution of \emph{ntpq}. We craft a prompt containing the stack trace and the error message as shown in ~\cref{fig:root_cause_analysis_1}, to discover the root cause of such a failure.

ChatGPT is able to identify the root cause perfectly in this case. It first identifies the vulnerability as stack smashing according to the stack trace, and then localizes the vulnerability at the line of code \emph{name[i] = cp}. In addition, though not requested, ChatGPT also provides possible fixes for the vulnerability. All the above results are correct based on our manual examination. 

\begin{figure}[t]
    \centering
    \includegraphics[width=0.49\textwidth]{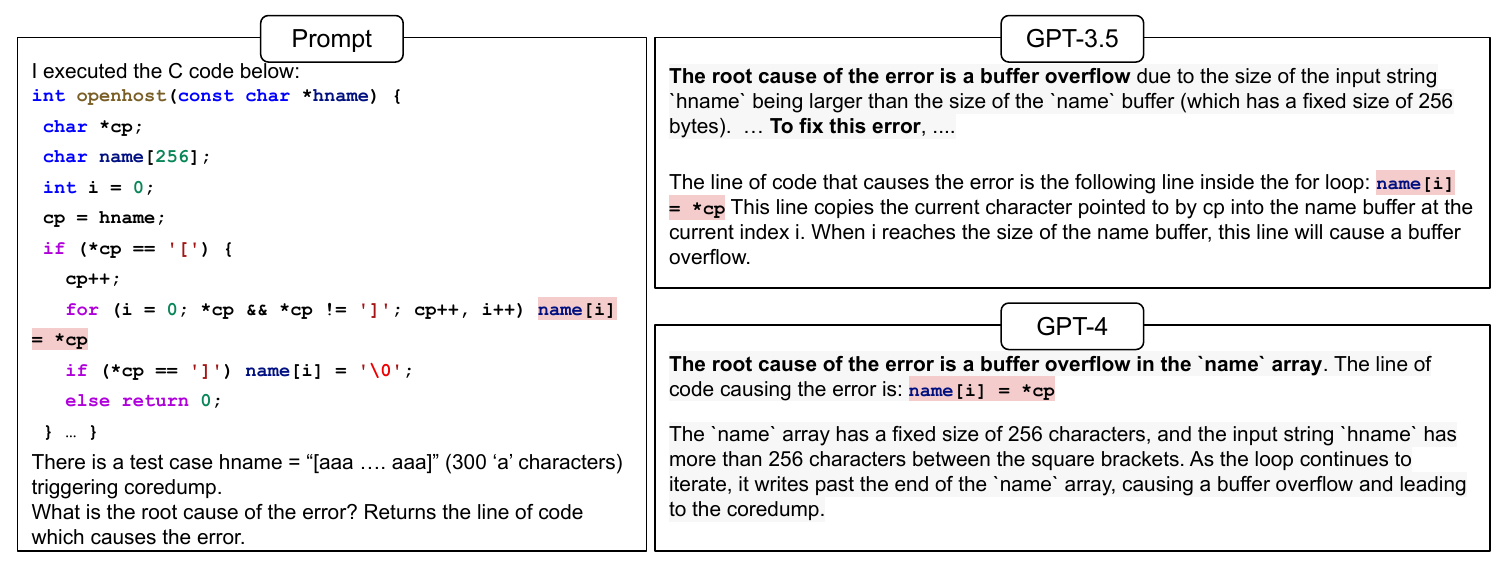}
    \caption{Given a failing test case, find the root cause of stack overflow in \emph{ntpq} (CVE-2018-12327).}
    \label{fig:root_cause_analysis_2}
\end{figure}

\textbf{Root cause analysis based on failed test cases}.
In this case study, we reuse the example of CVE-2018-12327 but change the hint of root cause analysis from error messages to a failing test case. As shown in ~\cref{fig:root_cause_analysis_2}, the failing test case triggers stack overflow by assigning \emph{hname} a very long name.
ChatGPT identifies the buffer overflow issue and localizes the critical code correctly. ChatGPT also provides reasonable explanations of the mechanism of the exploit.                      
\mysection{Decompilation}
\label{sec:decompliation}

Decompilation is the process of analyzing machine code and producing a high-level representation of the original source code. It is a reverse engineering technique that allows developers to understand how a program works by converting machine code back into a human-readable form.

Decompilation is an important technique for software security because it allows security professionals and researchers to analyze the behavior of a program and identify potential security vulnerabilities or malicious code that may be presented~\cite{reiter2022automatically,mantovani2022convergence,verbeek2020sound,schwartz2013native}. Through the process of decompiling an application project, security researchers can gain a deeper understanding of its functionality and pinpoint potential vulnerabilities that can be exploited.

However, the process of decompilation is not always straightforward due to various factors. Code obfuscation and anti-debugging techniques~\cite{branco2012scientific} can significantly complicate the decompilation process. Furthermore, the decompiled code may not always precisely match the original source code~\cite{schulte2018evolving}. Therefore, developers must rely on their understanding of the original code to interpret the decompiled output accurately.

LLMs now have the potential capability to understand the relationship between assembly code and the source code, which opens up a new avenue for decompilation that does not require execution environments. In this section, we analyze the capability of ChatGPT in decompilation, focusing on the translation from assembly code to source code. Binary-to-assembly translation is addressed by existing tools such as objdump and IDA Pro~\cite{ida} accurately. Therefore, ChatGPT can rely on the tools to preprocess the binary. Assambly-to-source translation is however challenging considering the rich semantics of source code.

\textbf{Dataset.} To assess ChatGPT's decompilation capabilities, we conducted an empirical analysis with real-world programs. A set of 30 programming solutions, drawn from LeetCode~\cite{leetcode} and written in the C programming language, served as our test base. These solutions were first compiled into binary code using gcc on the x86\_64 platform, and subsequently disassembled into assembly code via IDA Pro. The obtained assembly code constituted our benchmark, with the original C source code functioning as the ground truth. To ensure a balanced representation, the selected LeetCode problems span easy, medium, and hard levels of difficulty (10 problems in each difficulty level), and the source code length ranges from 10 to 70 lines.

\textbf{Prompts} We invoke ChatGPT with the prompt ``Decompile the following disassembly to C program'' and attach the assembly code produced by IDA Pro. We manually postprocess the responses from ChatGPT by extracting the decompiled source code. 

\textbf{Metrics.} We use the three metrics to measure how accurate the decompilation is.
(1) \emph{Name match.} Variable names are lost when the code is compiled to binary in release but decomplier could predict the variable names based on the code logic.
We count the matches between variable names in the orignal code and decompiled code, following DIRE~\cite{lacomis2019dire}.
(2) \emph{Type match.} Following~\cite{chen2022augmenting}, we assess the type prediction accuracy, which is similar to the name match. By comparing variable and function types between the original and decompiled code, we determine the number of variables and functions that have retained their types.
(3) \emph{Correctness.} To evaluate if the decompiled source code maintains the original functionality, we use the code to answer the corresponding LeetCode coding problem on the website and examine how many tests from LeetCode's test suites are passed. To be specific, the code may produce three outcomes for each test: passed, wrong results, or failed execution because of compiling problems or fatal errors.
In addition, we choose the well-known binary analysis tool IDA Pro~\cite{ida} to accomplish the same decompilation tasks and use the results as a baseline.

\begin{table}[t]
\scriptsize
\vspace{0.10in}
\caption{Decompilation on LeetCode solutions.}
\setlength{\tabcolsep}{2pt}
  \label{tab:decom}
  \centering
  \begin{tabular}{| c | c | c | c | c | c |}
    \noalign{\global\arrayrulewidth1pt}\hline\noalign{\global\arrayrulewidth0.4pt}
   \multirow{2}{*}{Tool} & \multirow{2}{*}{Name match} & \multirow{2}{*}{Type match} & \multicolumn{3}{c|}{Correctness} \\
   \cline{4-6}
   & & & Passed & Wrong results & Failed execution \\
    \hline
    IDA & 21.40\% & 42.47\% & 33.33\% &  {10\%} & \textbf{56.67\%} \\
    GPT-3.5 & 45.49\% & 59.20\% & 43.33\% & \textbf{26.67\%} & {30.00\%} \\
    GPT-4 & \textbf{56.52\%} & \textbf{78.26\% }& \textbf{53.33\%} & 20.00\% & 26.67\% \\
    \hline
    \noalign{\global\arrayrulewidth1pt}\hline\noalign{\global\arrayrulewidth0.4pt}
  \end{tabular}
\vspace{-0.20in}
\end{table}

\textbf{Disassembly-to-source-code translation}.
Table~\ref{tab:decom} showcases that ChatGPT has a superior capability on decompilation compared with IDA Pro on the LeetCode test data. 
GPT-4 correctly infers the correct name of 56.52\% variable identifiers, an enhancement of 11.03\% over GPT-3.5, and 35.12\% over IDA Pro. Likewise, GPT-4 demonstrates proficiency in predicting data types, with a correctness rate of 78.26\%, significantly outperforming IDA Pro by 35.79\%.
In terms of the functional correctness, code produced by GPT-4 passes 53.33\% of the test suites which is 10\% higher than GPT-3.5 and 20\% higher than IDA Pro. Notably, IDA Pro produces a substantial number of programs (56.67\%) that fail at the compilation stage, approximately 26\% more than GPT models. However, GPT models occasionally yield misleading programs, exhibiting incorrect execution results in 20-27\% of instances.

\begin{figure}[t]
    \centering
    \includegraphics[width=0.49\textwidth]{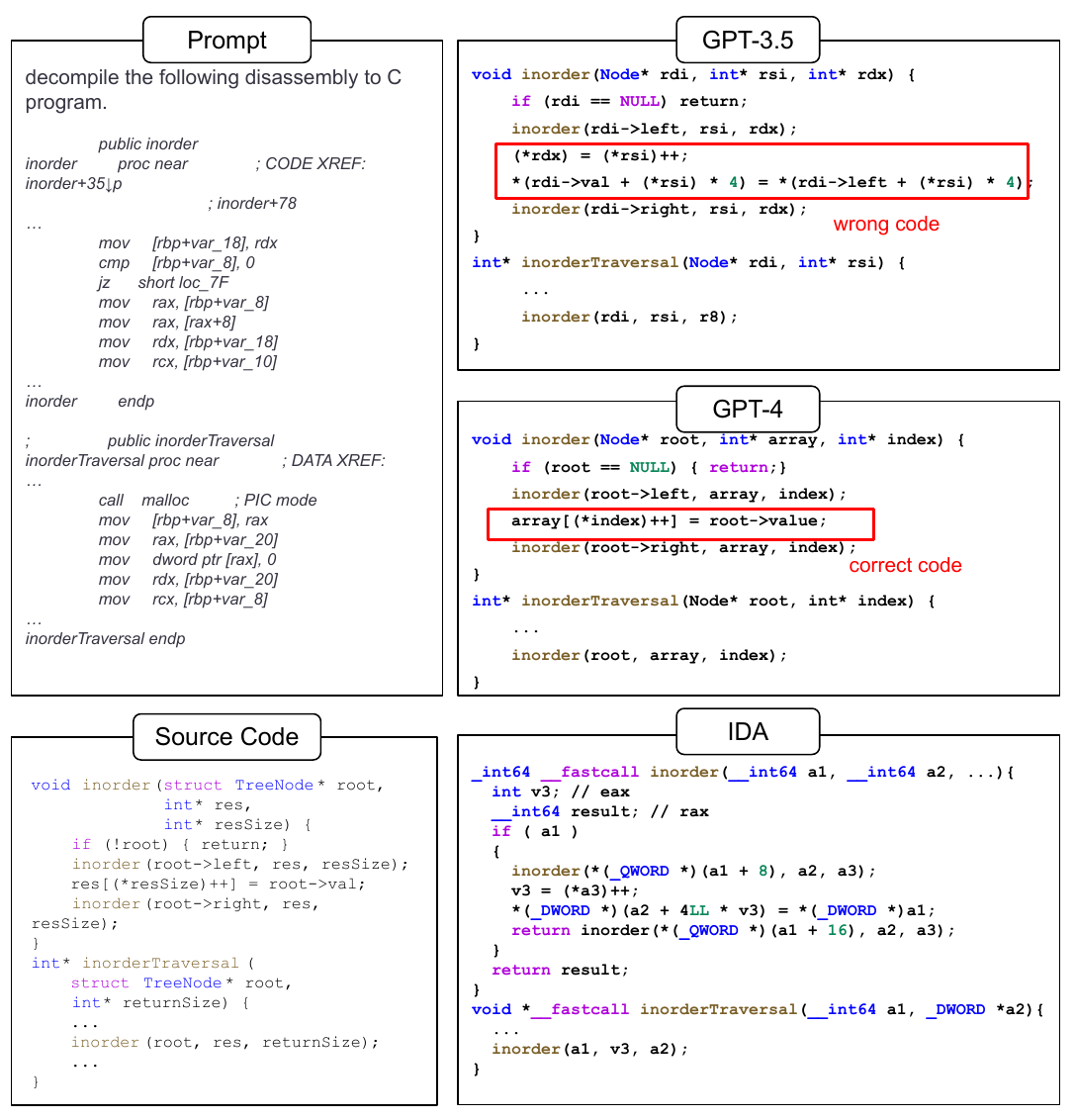}
    \caption{Ask ChatGPT to decompile an assembly code. Tested on a simple LeetCode C solution program and its x86\_64 assembly code.}
    \label{fig:decompilation_1}
\end{figure}

We then use a case study to demonstrate ChatGPT's decompilation capability.
As shown in ~\cref{fig:decompilation_1}, we ask ChatGPT to decompile a piece of assembly code back to C source code. 
The code is supposed to implement the algorithm of traversal a binary tree using recursion. The \emph{inorderTraversal} function is the entry of the algorithm while the \emph{inorder} function is recursively invoked during the code execution.
IDA Pro easily recovers the function names from the function references in the binary files and makes the overall code compilable and executable. IDA Pro's decompilation is in general preservative. It leverages only existing information from the binary and focuses on the correctness of code, instead of making the code similar to human-written code by inferring readable variable names and data types.
For instance, the variable names are meaningless such as \emph{v1}, and members of structs are represented by pointers and offsets.
On the other hand, ChatGPT has a strong capability to predict what the original code looks like. It is evident that ChatGPT infers the semantic of the code and assigns variables in code with frequently used names. For instance, GPT-4 accurately names the root of the binary tree with \emph{root} and correctly represents the two child nodes as \emph{root$\to$left} and \emph{root$\to$right}. Such informative naming significantly enhances the readability of the code, which is essential for advanced decompilation.
GPT-3.5 can also infer some of the variable names and types correctly, but makes a mistake on the code logic, resulting in wrong execution results.
GPT-4 demonstrates a major improvement from GPT-3.5 in terms of either prediction of names and correctness of code.


Overall, GPT models show impressive decompilation capabilities, surpassing the IDA Pro baseline across nearly all evaluated metrics on our LeetCode test cases. The results show that GPT's language processing capabilities extend beyond human languages and source code, demonstrating the ability to parse low-level assembly languages and comprehend complex underlying logic.

\textbf{Binary-to-disassembly translation}.
ChatGPT can hardly disassemble binary files to assembly code in our experiments. We design the prompt ``Disassemble the following binary to assembly code''. We feed the prompt as well as a hexadecimal string of a x86\_64 binary to ChatGPT. ChatGPT with either GPT-3.5 or GPT-4 cannot produce the disassembly. Instead, it provides a suggestion that the user should transform the hexadecimal string to a binary file in the operating system and use a disassembler or debugger to finish the job.
In general, binary files in hexadecimal string are much longer than the original source code and lack semantics, which is fundamentally challenging for LLMs to process.                           
\mysection{Debloating}
\label{sec:debloating}

Software debloating is the process of removing unnecessary code or features from a software application to reduce its attack surface and improve its security. For example, a web browser may have features such as certain plugins or extensions that can be used by attackers to compromise the user's system. Removing the unnecessary feature is a promising way to exhibit attacks fundamentally.

There are several approaches for software debloating, including static analysis~\cite{llvm,SQLite,EGLIBC,Toybox,BusyBox,jiang2016jred} and dynamic analysis~\cite{quach2018debloating,rastogi2017cimplifier,ghavamnia2020temporal,azad2019less}. 
Several studies~\cite{heo2018effective,ghaffarinia2019binary,xue2019hecate} also propose to leverage deep learning/machine learning models to identify code for removal in a learning-based fashion. Similarly, given ChatGPT's impressive ability to understand source code, it holds promise for conducting efficient debloating directly at the source-code level. In this section, we explore ChatGPT's potential for software debloating through both simple code examples and real-world scenarios.

\begin{figure}[t]
    \centering
    \includegraphics[width=0.49\textwidth]{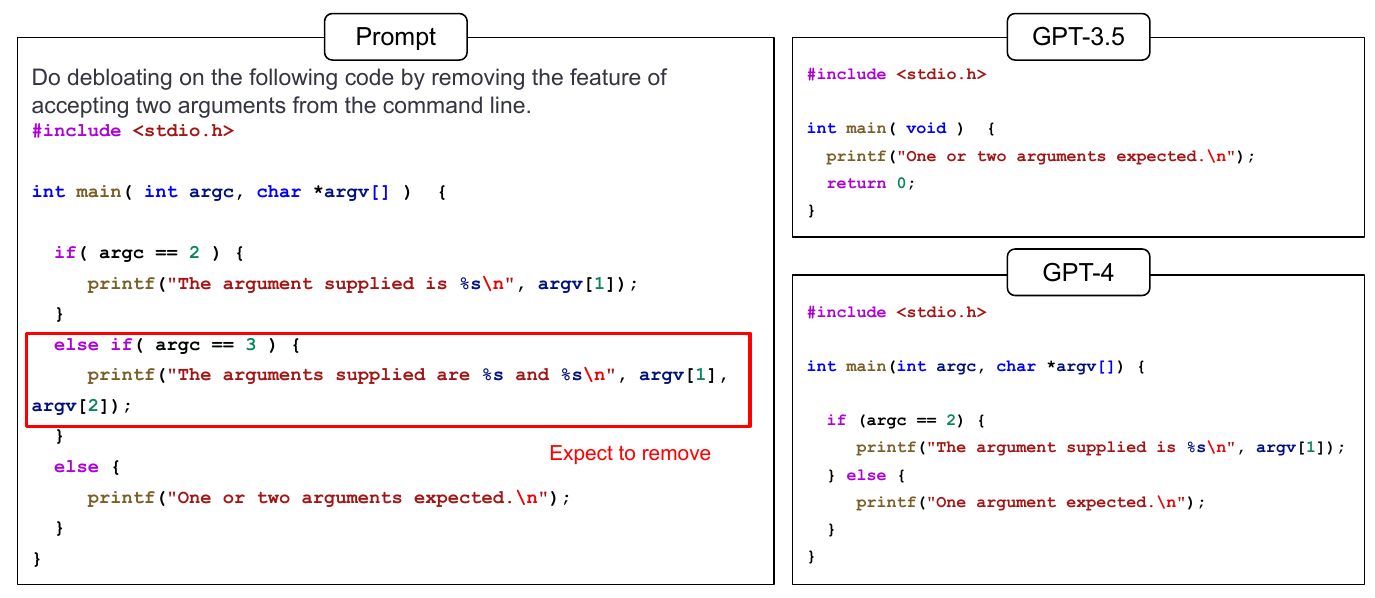}
    \caption{Ask ChatGPT to do debloating on a C program.}
    \label{fig:debloating_1}
\end{figure}

\textbf{Debloating simple programs}.
We initially debloat a simple C program, as illustrated in Figure~\ref{fig:debloating_1}. This program accepts command-line arguments and exhibits distinct behavior based on the number of arguments provided. It outputs the arguments when there are one or two, while displaying an error message otherwise. Our objective is to remove the functionality of accepting two arguments from the code. To accomplish this, we engage ChatGPT by presenting the source code along with the prompt ``Do debloating on the following code by removing the feature of accepting two arguments from the command line.''

ChatGPT understands the request and successfully removes part of the original code while keeping the remaining part untouched.
ChatGPT with GPT-4 gives a perfect solution which deletes the \emph{IF} code block of the two-argument scenario.
With GPT-3.5, ChatGPT initially fails to generate a valid solution as it mistakenly removes the one-argument feature. However, upon multiple attempts to regenerate responses, it successfully produces the correct answer in two subsequent tries. This outcome showcases ChatGPT's capability in software debloating, albeit with inaccuracies, particularly in the GPT-3.5 version.

\begin{figure}[t]
    \centering
    \includegraphics[width=0.49\textwidth]{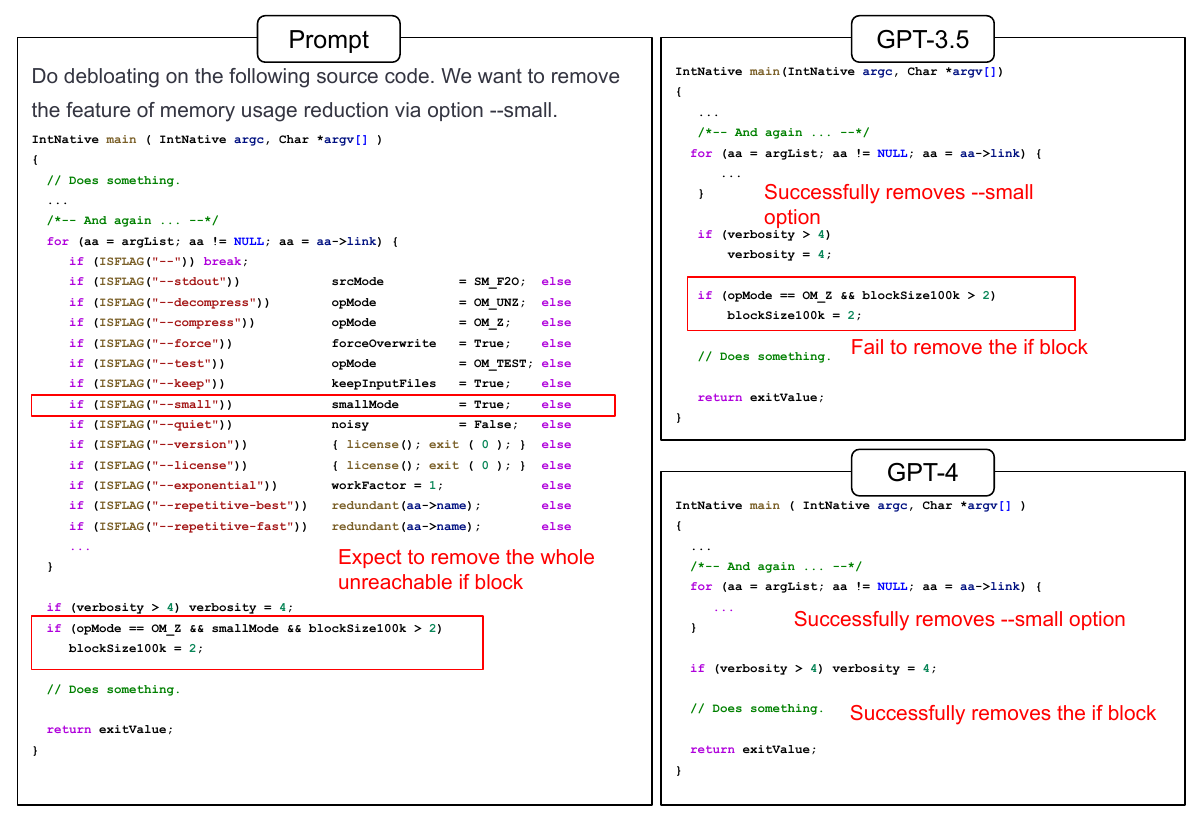}
    \caption{Ask ChatGPT to do debloating on part of the main function of \emph{bzip2}.}
    \label{fig:debloating_2}
\end{figure}

\textbf{Debloating real-world projects}.
We construct a debloating task using the source code from \emph{bzip2}~\cite{bzip2}. As shown in ~\cref{fig:debloating_2}, we aim to remove any source code related to feature of memory usage reduction, which is controlled by the command-line option \emph{--small} in \emph{bzip2}'s main function.

ChatGPT is capable of debloating on such real production code but inaccuracy exists. ChatGPT with either GPT-3.5 or GPT-4 removes the block about \emph{--small} in the \emph{FOR} loop.
Another variable \emph{smallMode} is related to the \emph{--small} option. If disabling \emph{--small} option, we should assume variable \emph{smallMode} as \emph{false} and also remove redundant code using \emph{smallMode}.
ChatGPT with GPT-4 handles \emph{smallMode} correctly by removing the \emph{IF} block whose condition is certain to be \emph{false}.
ChatGPT with GPT-3.5 attampts to remove \emph{smallMode} in the same \emph{IF} block but it assumes \emph{smallMode} to be \emph{true}, which is not correct.

It is worth nothing that it has a few challenges for using ChatGPT in real-world debloating tasks.
First, software debloating is usually for complex software that is non-trivial to trim but ChatGPT is limited in processing long content. Second, software debloating may work on binary files including linked libraries, which cannot be processed by ChatGPT originally.  
\mysection{Symbolic Execution}
\label{sec:symbolic_execution}

Symbolic execution~\cite{zhou2022ferry,boyer1975select,howden1977symbolic,king1975new,cadar2008klee,aschermann2019redqueen,chen2018angora,chen2019matryoshka,chipounov2011s2e} is a technique used in software engineering to analyze the behavior of a program by exploring all possible inputs and execution paths in a systematic manner.
In symbolic execution, a program is executed with symbolic inputs instead of actual values. These symbolic inputs are represented as variables with unknown values, called symbolic variables. As the program executes, the symbolic values of the variables are tracked and manipulated, allowing the analysis to explore all possible execution paths.
By exploring and exploiting possible execution paths, symbolic execution is promising to identify deeply hidden bugs and vulnerabilities as well as generate test cases that cover the vulnerable execution paths.

One limitation of symbolic execution is the path explosion~\cite{anand2008demand}. When the program grows larger and more complicated, the number of possible execution paths exponentially increases, causing an issue of scalability. Though various path pruning and selection methods~\cite{schwartz2015conflict,ma2011directed,kuznetsov2012efficient,poeplau2020symbolic} are designed to mitigate the path explosion problem, symbolic execution is still an expensive dynamic analysis method with heavy computation requirements.

However, ChatGPT is potential to replace certain expensive components in symbolic execution, such as solving the path conditions as Boolean satisfiability (SAT) problems, in order to efficiently explore more paths.
In this section, we demonstrate ChatGPT's capability on symbolic execution, especially the extraction of execution paths and SAT solving.

\begin{figure}[t]
    \centering
    \includegraphics[width=0.49\textwidth]{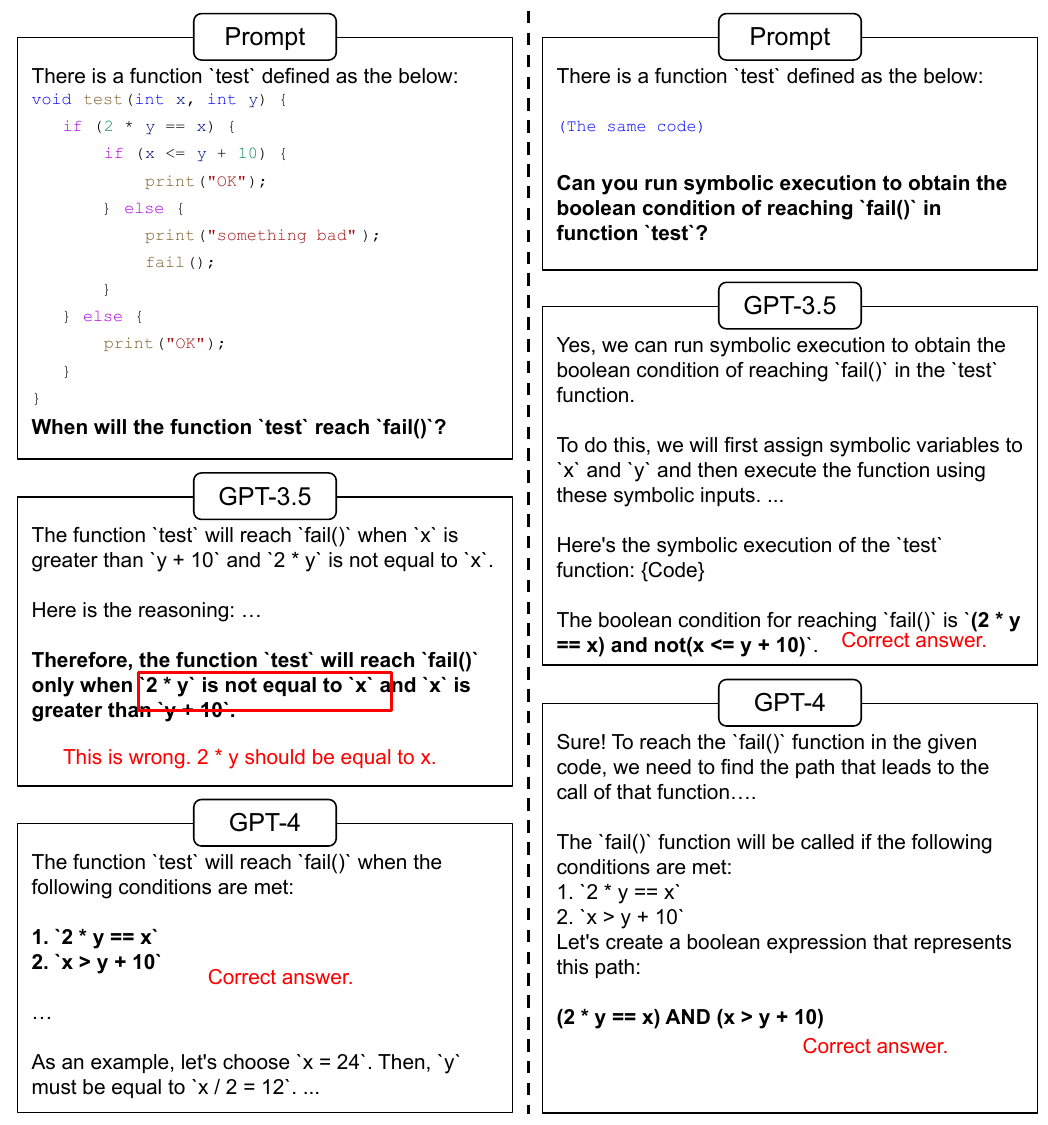}
    \caption{Solving symbolic execution problems on simple programs. ChatGPT is able to reason conditions in code while GPT-4's result is more concise and stable.}
    \label{fig:symbolic_execution_1}
\end{figure}

\textbf{Exploring execution paths in simple programs}.
In ~\cref{fig:symbolic_execution_1}, we demonstrate symbolic execution using ChatGPT on a simple C program. The C program defines a simple function that takes two integers as arguments, and one of the execution paths will reach a failure state. The user intends to understand in what situation the program will fail.

In the first prompt, we ask the question ``When will the function \emph{test} reach \emph{fail()}?'' In this case, ChatGPT is able to provide detailed logical reasoning such as extracting key \emph{IF} conditions in the code. It has the capability of basic Boolean operations, such as knowing the negation of ``x $\le$ y + 10'' is ``x $>$ y + 10''. However, ChatGPT does not guarantee the correctness of such logic reasoning. For instance, with GPT-3.5, ChatGPT gives a wrong conclusion saying ``2 * y is not equal to x'', which also contradicts its own explanation in the context.
ChatGPT with GPT-4 gives a good answer ``2 * y == x'' along with a concretized example ``x = 24, y = 12''.
In the second prompt, we write the prompt using a more formal presentation, asking ChatGPT to run ``symbolic execution'' and return ``boolean expression''. ChatGPT understands the request and produces correct boolean expressions representing the condition of the execution path to the failure state. GPT-4 answers ``(2 * y == x) AND (x $>$ y + 10)'', which is slightly conciser than CPT-3.5's answer ``(2 * y == x) and not(x $\le$ y + 10)''.

In summary, ChatGPT demonstrates impressive capability in reasoning program control flows, though wrong reasoning happens sometimes. Using proper prompts potentially make the responses conciser and more accurate.

\begin{figure}[t]
    \centering
    \includegraphics[width=0.49\textwidth]{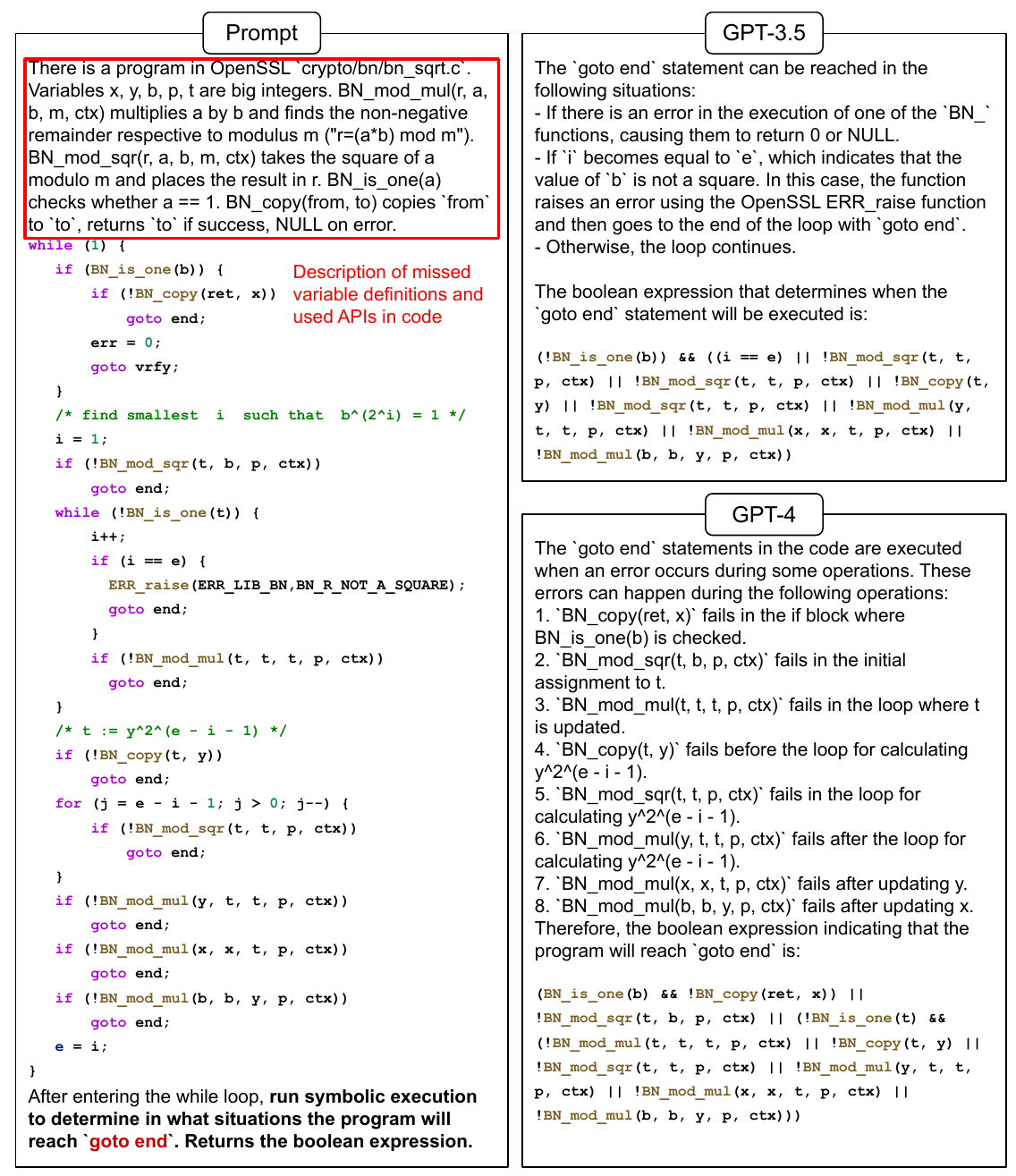}
    \caption{Solving symbolic execution problems on real-world software.}
    \label{fig:symbolic_execution_2}
\end{figure}

\textbf{Real-world scenarios of symbolic execution}.
To further evaluate ChatGPT's capability in real-world scenarios, we construct a symbolic execution task from CVE-2022-0778, a real vulnerability of OpenSSL. In this vulnerability case, if a non-prime modulus is passed in \emph{BN\_mod\_sqrt} as a variable, it triggers an infinite loop thus a denial-of-service (DoS). Since a special input triggers a specific vulnerable execution path, symbolic execution is a good fit to discover such vulnerability.

As shown in ~\cref{fig:symbolic_execution_2}, we ask ChatGPT for the condition on input variables that can make the loop (the given code) stop safely (\emph{goto end}). In addition, we add a description of input variables and APIs used in the code piece. Without the description, ChatGPT can hardly understand the whole code as the definitions of variables and function calls are not involved in the prompt.
ChatGPT can generally identify the key \emph{IF} conditions which are control dependents of the sink statements \emph{goto end}. However, when calculating the final Boolean expression, it simply concatenates these \emph{IF} conditions without considering the loop execution. For instance, inside the original \emph{WHILE}, there is another one \emph{WHILE} loop with local counter \emph{i} and one \emph{FOR} loop with local counter \emph{j}. Whether the code will reach \emph{goto end} also depends on the two smaller loops but ChatGPT does not provide insights on that.

From the case study, we find that ChatGPT originally is hard to resolve complicated control flows such as loops. Future improvements or system support is needed to integrate ChatGPT in large-scale real-world symbolic execution.                      
\mysection{Fuzzing}
\label{sec:fuzzing}

Fuzzing is an effective software testing technique that entails supplying programs with invalid, unexpected, or random data inputs to identify vulnerabilities and bugs. The fundamental concept behind fuzzing is to stress-test a program by inundating it with a large volume of diverse input data, thereby uncovering defects that may remain undetected by conventional testing approaches.

The technique of fuzzing could be categorized into two classes: mutation-based and generation-based. Mutation-based fuzzing~\cite{zhu2022fuzzing,manes2019art,peng2018t,fayaz2016buzz,rawat2017vuzzer} involves taking existing input data and making small, random changes to it to create new test cases. By scheduling seed test cases and mutation operators, such fuzzing tools aim to produce interesting test cases efficiently. Generation-based fuzzing~\cite{zhu2022fuzzing,manes2019art,kaksonen2001software,dharma,vuagnoux2005autodafe,abdelnur2007kif} is aware of the structure of the valid test cases. This is very useful when the subject software only accepts data in certain formats, such as PDF readers and compilers. Generation-based fuzzing effectively avoids test cases in invalid formats thus saving time for more useful tests.
In addition, there are also hybrid fuzzers~\cite{zhu2022fuzzing,manes2019art,stephens2016driller,yun2018qsym} that leverage multiple fuzzing techniques.

Existing fuzzing tools demonstrate great success in discovering software vulnerabilities. AFL~\cite{afl}, AFL++~\cite{afl++}, libfuzzer~\cite{libfuzzer} are examples of the most popular automatic fuzzing tools, which have revealed thousands of software vulnerabilities in widely used projects~\cite{serebryany2017oss}.
Previous research keeps improving fuzzing by designing intelligent scheduling of seeds or mutation, aiming to efficiently cover buggy or vulnerable execution paths in the large code base~\cite{metzman2021fuzzbench}. Recently, researchers find that LLMs can assist fuzzing process using its strong ability of generating source code. TitanFuzz~\cite{deng2022fuzzing}, for instance, use a LLM to generate deep-learning related code to reveal vulnerabilities in deep learning libraries such as Torch~\cite{collobert2002torch} and Tensorflow~\cite{abadi2016tensorflow}. ChatGPT as one of the state-of-art LLMs also has the potential to tackle with fuzzing challenges.
In this section, we study ChatGPT's capability in fuzzing involving both mutation-based and generation-based fuzzing scenarios.

\textbf{Mutation-based fuzzing}.
In conventional mutation-based fuzzing, a set of mutation operators are pre-defined and a set of seed test cases is updated at runtime.
For instance, if the seed test case is a string ``Hello, world!'', the fuzzing process may change one byte in the string to produce new test cases like ``Hell@, world!''.
However, the implementation of the mutation could be expensive in human effort. ChatGPT may be able to automate such a process given a simple description.

\begin{figure}[t]
    \centering
    \includegraphics[width=0.49\textwidth]{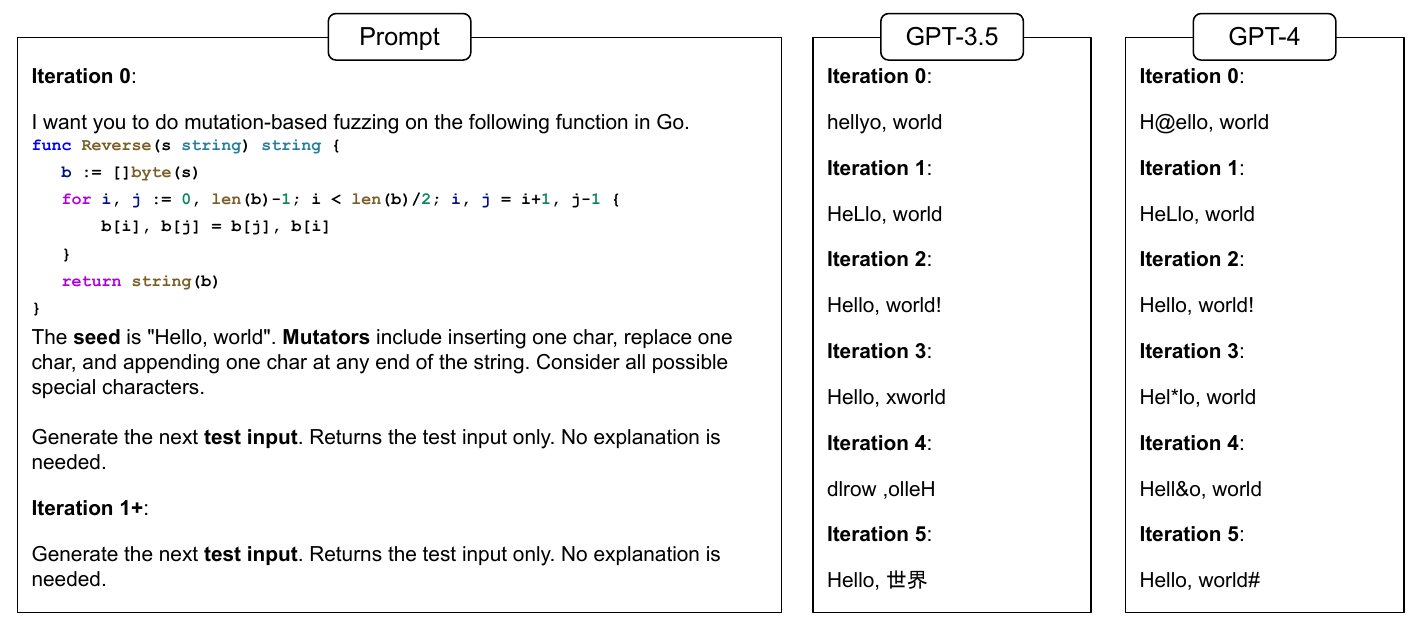}
    \caption{Make ChatGPT performs like a mutation-based fuzzer. Tested on a simple Go function.}
    \label{fig:fuzzing_1}
\end{figure}

As shown in ~\cref{fig:fuzzing_1}, we provide ChatGPT with the source code to test, a predefined seed test case, and a set of mutation operators. Then we ask ChatGPT to generate a test case for the source code. After obtaining responses from ChatGPT, we repeat the request of test case generation for multiple iterations, similar to how a fuzzer works.

From the results, ChatGPT with either GPT-3.5 or GPT-4 has sufficient capability to perform the fuzzing process. It can understand the concept of mutation-based fuzzing, including ``seeds'' and ``motators''. ChatGPT is also aware of the meaning of mutations, such as byte manipulations.
For example, when the seed is ``Hello, world'', ChatGPT can produce test cases such as ``hellyo, world'' by injecting bytes and ``HeLlo, world'' by replacing a byte.

However, ChatGPT can hardly enforce hard constraints on the mutation process, which means it may apply mutations that are out of the definition of mutation operators. For instance, ChatGPT with GPT-3.5 produces a test case by replacing ``world'' with its Chinese translation, which is out of the pre-defined set of mutators.
In contrast, ChatGPT with GPT-4 obeys the mutation rules as far as we see in the example, demonstrating a stronger capability of understanding mutation-related requirements.

\textbf{Generation-based fuzzing}.
Generation-based fuzzing emphasizes the capability of enforcing generated test cases to be compliant with certain data formats or grammar rules. Conventionally, this is implemented by using a pre-defined specification or grammar. ChatGPT is built on LLM which is trained on the vast amount of corpus thus it has the potential to understand complex grammar.

\begin{figure}[t]
    \centering
    \includegraphics[width=0.49\textwidth]{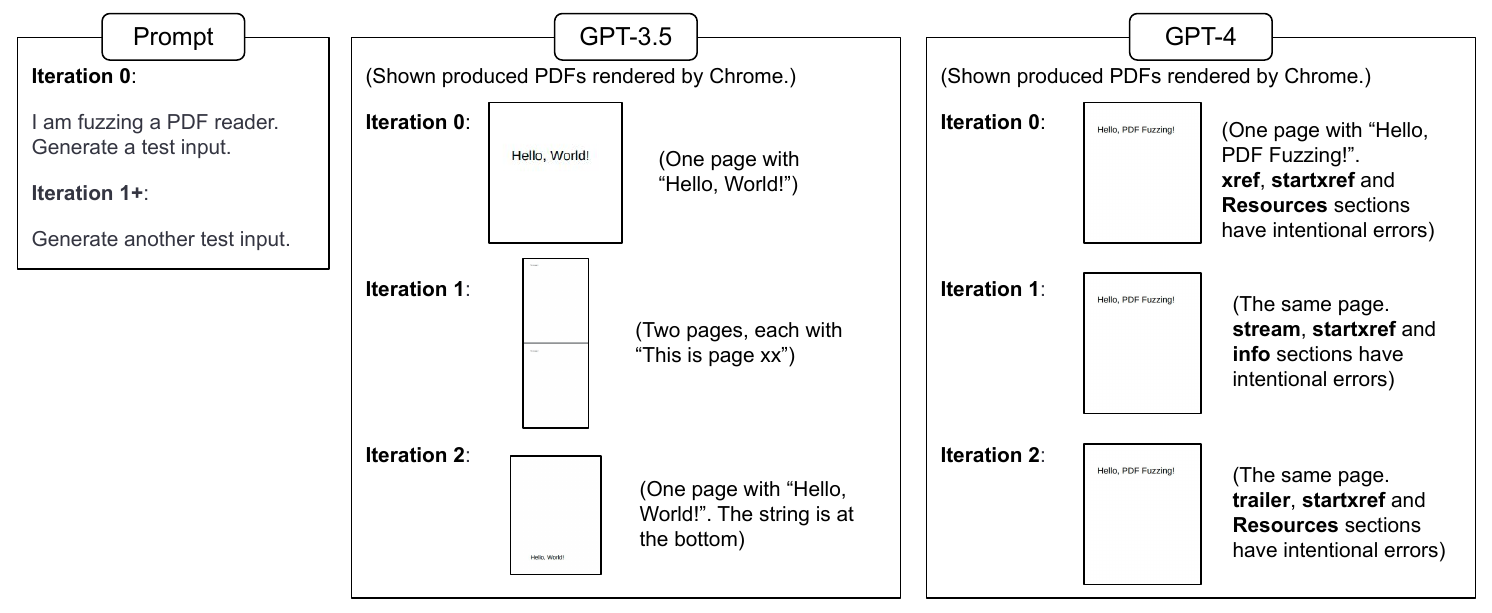}
    \caption{Make ChatGPT performs like a generation-based fuzzer that tests PDF readers.}
    \label{fig:fuzzing_2}
\end{figure}

As shown in ~\cref{fig:fuzzing_2}, we ask ChatGPT to perform fuzzing on a PDF reader, expecting it to produce PDF files that may trigger internal errors in the PDF reader.
We repeat the request to obtain multiple generated test cases.

ChatGPT with either GPT-3.5 or GPT-4 can generate valid PDF test cases. 
However, GPT-3.5 and GPT-4 seem to focus on different aspects in terms of the diversity of PDF files.
ChatGPT with GPT-3.5 generates PDF files with different appearance, e.g., changing the location of text blocks or increasing the number of pages.
Instead, ChatGPT with GPT-4 intends to generate PDF files with errors in certain sections, such as false entries in the ``trailer'' dictionary and wrong ``startref'' value. GPT-4 also provides a detailed description of the errors it injected.
In comparison, generated test cases from GPT-4 are more valuable for fuzzing as they contain intentional errors that may fail PDF parsing. GPT-4 in general demonstrates a stronger understanding of the security context of fuzzing.

\textbf{Towards closed loop fuzzing}.
From the above results, ChatGPT especially the GPT-4 version demonstrates a strong capability in generating test cases for fuzzing. However, fuzzing is a closed loop process involving the feedback from the execution environment, the optimization of seed scheduling based on the feedback, and other system-level challenges. Though powerful, ChatGPT as a text processor cannot fabricate the whole fuzzing process alone. It is a future research direction to integrate concrete system components with the generation capability of ChatGPT.                                 
\mysection{Conclusion and Future Work}
\label{sec:conclusion}

Based on extensive experiments and thorough case studies, it is clear that ChatGPT, especially when powered by the remarkable GPT-4, exhibits exceptional potential in assisting with a wide range of software security tasks. 
These tasks include but are not limited to vulnerability detection, vulnerability repair, bug fixing, patching, debloating, decompiling, root cause analysis, symbolic execution, and fuzzing. Notably, ChatGPT and GPT models demonstrate strong proficiency in processing source code, and they even exhibit promising capabilities in processing disassembly code. Moreover, ChatGPT excels in human interaction, adapting its responses based on user-provided task descriptions.
The impressive performance of GPT-4 surpasses that of GPT-3.5 across most software security tasks, highlighting the continual advancements made in large language models.

An intriguing finding is ChatGPT's ability to handle complex data beyond traditional source code. In the task of decompilation, ChatGPT showcases the capacity to accurately transform disassembly code back into C source code, surpassing the capability of IDA Pro. Additionally, in the domain of generation-based fuzzing, ChatGPT successfully generates random benign or buggy PDF files, which can be correctly rendered and displayed. These findings provide strong motivation for future security research to embrace the integration of LLMs in more intricate tasks.

It is worth noting that while ChatGPT shows promise, its performance still falls short compared to established conventional methods in scalability, especially when dealing with large data volumes. The resource comsuption of LLMs increases significantly as the size of input increases, which make it difficult to handle software projects with at least a few thousand lines of code. Real-world security challenges often involve system-level information, such as computer architecture and network specifics, which presents difficulties in conveying such information effectively through textual sequences to ChatGPT. Additionally, ChatGPT lacks the ability to comprehend binary code, further contributing to its limitations. Nevertheless, addressing these challenges through future research holds significant promise.

While our study demonstrates valuable insights, we acknowledge potential threats to validity. Due to the absence of benchmark datasets, we focused on simple yet representative cases for our case studies, except for vulnerability detection/repair and bug fixing. Future work should extend these case studies to encompass comprehensive evaluations using larger and more diverse benchmark datasets.                              

%
\IEEEpeerreviewmaketitle

\bibliographystyle{IEEEtran}
\bibliography{IEEEabrv,ref}

\begin{thebibliography}{100}
\providecommand{\url}[1]{#1}
\csname url@samestyle\endcsname
\providecommand{\newblock}{\relax}
\providecommand{\bibinfo}[2]{#2}
\providecommand{\BIBentrySTDinterwordspacing}{\spaceskip=0pt\relax}
\providecommand{\BIBentryALTinterwordstretchfactor}{4}
\providecommand{\BIBentryALTinterwordspacing}{\spaceskip=\fontdimen2\font plus
\BIBentryALTinterwordstretchfactor\fontdimen3\font minus
  \fontdimen4\font\relax}
\providecommand{\BIBforeignlanguage}[2]{{%
\expandafter\ifx\csname l@#1\endcsname\relax
\typeout{** WARNING: IEEEtran.bst: No hyphenation pattern has been}%
\typeout{** loaded for the language `#1'. Using the pattern for}%
\typeout{** the default language instead.}%
\else
\language=\csname l@#1\endcsname
\fi
#2}}
\providecommand{\BIBdecl}{\relax}
\BIBdecl

\bibitem{rumelhart1986learning}
D.~E. Rumelhart, G.~E. Hinton, and R.~J. Williams, ``Learning representations
  by back-propagating errors,'' \emph{nature}, vol. 323, no. 6088, pp.
  533--536, 1986.

\bibitem{rumelhart1985learning}
------, ``Learning internal representations by error propagation,'' California
  Univ San Diego La Jolla Inst for Cognitive Science, Tech. Rep., 1985.

\bibitem{jordan1997serial}
M.~I. Jordan, ``Serial order: A parallel distributed processing approach,'' in
  \emph{Advances in psychology}.\hskip 1em plus 0.5em minus 0.4em\relax
  Elsevier, 1997, vol. 121, pp. 471--495.

\bibitem{hochreiter1997long}
S.~Hochreiter and J.~Schmidhuber, ``Long short-term memory,'' \emph{Neural
  computation}, vol.~9, no.~8, pp. 1735--1780, 1997.

\bibitem{vaswani2017attention}
A.~Vaswani, N.~Shazeer, N.~Parmar, J.~Uszkoreit, L.~Jones, A.~N. Gomez,
  {\L}.~Kaiser, and I.~Polosukhin, ``Attention is all you need,''
  \emph{Advances in neural information processing systems}, vol.~30, 2017.

\bibitem{radford2018improving}
A.~Radford, K.~Narasimhan, T.~Salimans, I.~Sutskever \emph{et~al.}, ``Improving
  language understanding by generative pre-training,'' 2018.

\bibitem{radford2019language}
A.~Radford, J.~Wu, R.~Child, D.~Luan, D.~Amodei, I.~Sutskever \emph{et~al.},
  ``Language models are unsupervised multitask learners,'' \emph{OpenAI blog},
  vol.~1, no.~8, p.~9, 2019.

\bibitem{brown2020language}
T.~Brown, B.~Mann, N.~Ryder, M.~Subbiah, J.~D. Kaplan, P.~Dhariwal,
  A.~Neelakantan, P.~Shyam, G.~Sastry, A.~Askell \emph{et~al.}, ``Language
  models are few-shot learners,'' \emph{Advances in neural information
  processing systems}, vol.~33, pp. 1877--1901, 2020.

\bibitem{gpt-4}
OpenAI, ``Gpt-4 technical report,'' 2023.

\bibitem{chatgpt}
\BIBentryALTinterwordspacing
``Introducing chatgpt - open ai,'' 2023. [Online]. Available:
  \url{https://openai.com/blog/chatgpt}
\BIBentrySTDinterwordspacing

\bibitem{aljanabi2023chatgpt}
M.~Aljanabi, M.~Ghazi, A.~H. Ali, S.~A. Abed \emph{et~al.}, ``Chatgpt: Open
  possibilities,'' \emph{Iraqi Journal For Computer Science and Mathematics},
  vol.~4, no.~1, pp. 62--64, 2023.

\bibitem{gozalo2023chatgpt}
R.~Gozalo-Brizuela and E.~C. Garrido-Merchan, ``Chatgpt is not all you need. a
  state of the art review of large generative ai models,'' \emph{arXiv preprint
  arXiv:2301.04655}, 2023.

\bibitem{liu2023summary}
Y.~Liu, T.~Han, S.~Ma, J.~Zhang, Y.~Yang, J.~Tian, H.~He, A.~Li, M.~He, Z.~Liu
  \emph{et~al.}, ``Summary of chatgpt/gpt-4 research and perspective towards
  the future of large language models,'' \emph{arXiv preprint
  arXiv:2304.01852}, 2023.

\bibitem{bubeck2023sparks}
S.~Bubeck, V.~Chandrasekaran, R.~Eldan, J.~Gehrke, E.~Horvitz, E.~Kamar,
  P.~Lee, Y.~T. Lee, Y.~Li, S.~Lundberg \emph{et~al.}, ``Sparks of artificial
  general intelligence: Early experiments with gpt-4,'' \emph{arXiv preprint
  arXiv:2303.12712}, 2023.

\bibitem{azaria2022chatgpt}
A.~Azaria, ``Chatgpt usage and limitations,'' 2022.

\bibitem{maddigan2023chat2vis}
P.~Maddigan and T.~Susnjak, ``Chat2vis: Generating data visualisations via
  natural language using chatgpt, codex and gpt-3 large language models,''
  \emph{IEEE Access}, 2023.

\bibitem{frieder2023mathematical}
S.~Frieder, L.~Pinchetti, R.-R. Griffiths, T.~Salvatori, T.~Lukasiewicz, P.~C.
  Petersen, A.~Chevalier, and J.~Berner, ``Mathematical capabilities of
  chatgpt,'' \emph{arXiv preprint arXiv:2301.13867}, 2023.

\bibitem{biswas2023role}
S.~Biswas, ``Role of chatgpt in gaming: According to chatgpt,'' \emph{Available
  at SSRN 4375510}, 2023.

\bibitem{tsai2023can}
C.~F. Tsai, X.~Zhou, S.~S. Liu, J.~Li, M.~Yu, and H.~Mei, ``Can large language
  models play text games well? current state-of-the-art and open questions,''
  \emph{arXiv preprint arXiv:2304.02868}, 2023.

\bibitem{sallam2023chatgpt}
M.~Sallam, ``Chatgpt utility in healthcare education, research, and practice:
  systematic review on the promising perspectives and valid concerns,'' in
  \emph{Healthcare}, vol.~11, no.~6.\hskip 1em plus 0.5em minus 0.4em\relax
  MDPI, 2023, p. 887.

\bibitem{zheng2023can}
M.~Zheng, X.~Su, S.~You, F.~Wang, C.~Qian, C.~Xu, and S.~Albanie, ``Can gpt-4
  perform neural architecture search?'' \emph{arXiv preprint arXiv:2304.10970},
  2023.

\bibitem{choi2023chatgpt}
J.~H. Choi, K.~E. Hickman, A.~Monahan, and D.~Schwarcz, ``Chatgpt goes to law
  school,'' \emph{Available at SSRN}, 2023.

\bibitem{ryznar2023exams}
M.~Ryznar, ``Exams in the time of chatgpt,'' \emph{Washington and Lee Law
  Review Online}, vol.~80, no.~5, p. 305, 2023.

\bibitem{newton2023chatgpt}
P.~M. Newton, ``Chatgpt performance on mcq-based exams,'' 2023.

\bibitem{bang2023multitask}
Y.~Bang, S.~Cahyawijaya, N.~Lee, W.~Dai, D.~Su, B.~Wilie, H.~Lovenia, Z.~Ji,
  T.~Yu, W.~Chung \emph{et~al.}, ``A multitask, multilingual, multimodal
  evaluation of chatgpt on reasoning, hallucination, and interactivity,''
  \emph{arXiv preprint arXiv:2302.04023}, 2023.

\bibitem{borji2023categorical}
A.~Borji, ``A categorical archive of chatgpt failures,'' \emph{arXiv preprint
  arXiv:2302.03494}, 2023.

\bibitem{dong2023self}
Y.~Dong, X.~Jiang, Z.~Jin, and G.~Li, ``Self-collaboration code generation via
  chatgpt,'' \emph{arXiv preprint arXiv:2304.07590}, 2023.

\bibitem{tian2023chatgpt}
H.~Tian, W.~Lu, T.~O. Li, X.~Tang, S.-C. Cheung, J.~Klein, and T.~F.
  Bissyand{\'e}, ``Is chatgpt the ultimate programming assistant--how far is
  it?'' \emph{arXiv preprint arXiv:2304.11938}, 2023.

\bibitem{liu2023your}
J.~Liu, C.~S. Xia, Y.~Wang, and L.~Zhang, ``Is your code generated by chatgpt
  really correct? rigorous evaluation of large language models for code
  generation,'' \emph{arXiv preprint arXiv:2305.01210}, 2023.

\bibitem{khoury2023secure}
R.~Khoury, A.~R. Avila, J.~Brunelle, and B.~M. Camara, ``How secure is code
  generated by chatgpt?'' \emph{arXiv preprint arXiv:2304.09655}, 2023.

\bibitem{pearce2022asleep}
H.~Pearce, B.~Ahmad, B.~Tan, B.~Dolan-Gavitt, and R.~Karri, ``Asleep at the
  keyboard? assessing the security of github copilot’s code contributions,''
  in \emph{2022 IEEE Symposium on Security and Privacy (SP)}.\hskip 1em plus
  0.5em minus 0.4em\relax IEEE, 2022, pp. 754--768.

\bibitem{pearce2022examining}
H.~Pearce, B.~Tan, B.~Ahmad, R.~Karri, and B.~Dolan-Gavitt, ``Examining
  zero-shot vulnerability repair with large language models,'' in \emph{2023
  IEEE Symposium on Security and Privacy (SP)}.\hskip 1em plus 0.5em minus
  0.4em\relax IEEE Computer Society, 2022, pp. 1--18.

\bibitem{sobania2023analysis}
D.~Sobania, M.~Briesch, C.~Hanna, and J.~Petke, ``An analysis of the automatic
  bug fixing performance of chatgpt,'' \emph{arXiv preprint arXiv:2301.08653},
  2023.

\bibitem{xia2023conversational}
C.~S. Xia and L.~Zhang, ``Conversational automated program repair,''
  \emph{arXiv preprint arXiv:2301.13246}, 2023.

\bibitem{deng2022fuzzing}
Y.~Deng, C.~S. Xia, H.~Peng, C.~Yang, and L.~Zhang, ``Fuzzing deep-learning
  libraries via large language models,'' \emph{arXiv preprint
  arXiv:2212.14834}, 2022.

\bibitem{ouyang2022training}
L.~Ouyang, J.~Wu, X.~Jiang, D.~Almeida, C.~Wainwright, P.~Mishkin, C.~Zhang,
  S.~Agarwal, K.~Slama, A.~Ray \emph{et~al.}, ``Training language models to
  follow instructions with human feedback,'' \emph{Advances in Neural
  Information Processing Systems}, vol.~35, pp. 27\,730--27\,744, 2022.

\bibitem{christiano2017deep}
P.~F. Christiano, J.~Leike, T.~Brown, M.~Martic, S.~Legg, and D.~Amodei, ``Deep
  reinforcement learning from human preferences,'' \emph{Advances in neural
  information processing systems}, vol.~30, 2017.

\bibitem{cheshkov2023evaluation}
A.~Cheshkov, P.~Zadorozhny, and R.~Levichev, ``Evaluation of chatgpt model for
  vulnerability detection,'' 2023.

\bibitem{copilot}
``Github copilot · your ai pair programmer,''
  \url{https://copilot.github.com/}, 2023.

\bibitem{yang2023mm}
Z.~Yang, L.~Li, J.~Wang, K.~Lin, E.~Azarnasab, F.~Ahmed, Z.~Liu, C.~Liu,
  M.~Zeng, and L.~Wang, ``Mm-react: Prompting chatgpt for multimodal reasoning
  and action,'' \emph{arXiv preprint arXiv:2303.11381}, 2023.

\bibitem{wu2023visual}
C.~Wu, S.~Yin, W.~Qi, X.~Wang, Z.~Tang, and N.~Duan, ``Visual chatgpt: Talking,
  drawing and editing with visual foundation models,'' \emph{arXiv preprint
  arXiv:2303.04671}, 2023.

\bibitem{chen2023gptutor}
E.~Chen, R.~Huang, H.-S. Chen, Y.-H. Tseng, and L.-Y. Li, ``Gptutor: a
  chatgpt-powered programming tool for code explanation,'' \emph{arXiv preprint
  arXiv:2305.01863}, 2023.

\bibitem{park2023generative}
J.~S. Park, J.~C. O'Brien, C.~J. Cai, M.~R. Morris, P.~Liang, and M.~S.
  Bernstein, ``Generative agents: Interactive simulacra of human behavior,''
  \emph{arXiv preprint arXiv:2304.03442}, 2023.

\bibitem{feng2023investigating}
Y.~Feng, S.~Vanam, M.~Cherukupally, W.~Zheng, M.~Qiu, and H.~Chen,
  ``Investigating code generation performance of chat-gpt with crowdsourcing
  social data,'' in \emph{Proceedings of the 47th IEEE Computer Software and
  Applications Conference}, 2023, pp. 1--10.

\bibitem{liu2023improving}
C.~Liu, X.~Bao, H.~Zhang, N.~Zhang, H.~Hu, X.~Zhang, and M.~Yan, ``Improving
  chatgpt prompt for code generation,'' \emph{arXiv preprint arXiv:2305.08360},
  2023.

\bibitem{baidoo2023education}
D.~Baidoo-Anu and L.~Owusu~Ansah, ``Education in the era of generative
  artificial intelligence (ai): Understanding the potential benefits of chatgpt
  in promoting teaching and learning,'' \emph{Available at SSRN 4337484}, 2023.

\bibitem{biswas2023potential}
S.~S. Biswas, ``Potential use of chat gpt in global warming,'' \emph{Annals of
  biomedical engineering}, vol.~51, no.~6, pp. 1126--1127, 2023.

\bibitem{xia2023automated}
C.~S. Xia, Y.~Wei, and L.~Zhang, ``Automated program repair in the era of large
  pre-trained language models,'' in \emph{Proceedings of the 45th International
  Conference on Software Engineering (ICSE 2023). Association for Computing
  Machinery}, 2023.

\bibitem{xia2023keep}
C.~S. Xia and L.~Zhang, ``Keep the conversation going: Fixing 162 out of 337
  bugs for \$0.42 each using chatgpt,'' \emph{arXiv preprint arXiv:2304.00385},
  2023.

\bibitem{jin2023inferfix}
M.~Jin, S.~Shahriar, M.~Tufano, X.~Shi, S.~Lu, N.~Sundaresan, and
  A.~Svyatkovskiy, ``Inferfix: End-to-end program repair with llms,'' 2023.

\bibitem{deng2023large}
Y.~Deng, C.~S. Xia, C.~Yang, S.~D. Zhang, S.~Yang, and L.~Zhang, ``Large
  language models are edge-case fuzzers: Testing deep learning libraries via
  fuzzgpt,'' 2023.

\bibitem{hu2023augmenting}
J.~Hu, Q.~Zhang, and H.~Yin, ``Augmenting greybox fuzzing with generative ai,''
  2023.

\bibitem{ko1994automated}
C.~Ko, G.~Fink, and K.~Levitt, ``Automated detection of vulnerabilities in
  privileged programs by execution monitoring,'' in \emph{Tenth Annual Computer
  Security Applications Conference}.\hskip 1em plus 0.5em minus 0.4em\relax
  IEEE, 1994, pp. 134--144.

\bibitem{xiao2020mvp}
Y.~Xiao, B.~Chen, C.~Yu, Z.~Xu, Z.~Yuan, F.~Li, B.~Liu, Y.~Liu, W.~Huo, W.~Zou
  \emph{et~al.}, ``Mvp: Detecting vulnerabilities using patch-enhanced
  vulnerability signatures.'' in \emph{USENIX Security Symposium}, 2020, pp.
  1165--1182.

\bibitem{jeong2019razzer}
D.~R. Jeong, K.~Kim, B.~Shivakumar, B.~Lee, and I.~Shin, ``Razzer: Finding
  kernel race bugs through fuzzing,'' in \emph{2019 IEEE Symposium on Security
  and Privacy (SP)}.\hskip 1em plus 0.5em minus 0.4em\relax IEEE, 2019, pp.
  754--768.

\bibitem{codeql}
``{CodeQL Documentation},'' \url{https://codeql.github.com/docs/}, 2023.

\bibitem{infer}
``{Infer static analyzer},'' \url{https://fbinfer.com/docs}, 2023.

\bibitem{clang}
``{Clang static analyzer},'' \url{https://clang-analyzer.llvm.org/}, 2023.

\bibitem{hu2021automated}
S.~Hu, Q.~A. Chen, J.~Sun, Y.~Feng, Z.~M. Mao, and H.~X. Liu, ``Automated
  discovery of denial-of-service vulnerabilities in connected vehicle
  protocols.'' in \emph{USENIX Security Symposium}, 2021, pp. 3219--3236.

\bibitem{li2020automated}
X.~Li, L.~Wang, Y.~Xin, Y.~Yang, and Y.~Chen, ``Automated vulnerability
  detection in source code using minimum intermediate representation
  learning,'' \emph{Applied Sciences}, vol.~10, no.~5, p. 1692, 2020.

\bibitem{russell2018automated}
R.~Russell, L.~Kim, L.~Hamilton, T.~Lazovich, J.~Harer, O.~Ozdemir,
  P.~Ellingwood, and M.~McConley, ``Automated vulnerability detection in source
  code using deep representation learning,'' in \emph{2018 17th IEEE
  international conference on machine learning and applications (ICMLA)}.\hskip
  1em plus 0.5em minus 0.4em\relax IEEE, 2018, pp. 757--762.

\bibitem{zagane2020deep}
M.~Zagane, M.~K. Abdi, and M.~Alenezi, ``Deep learning for software
  vulnerabilities detection using code metrics,'' \emph{IEEE Access}, vol.~8,
  pp. 74\,562--74\,570, 2020.

\bibitem{li2021automated}
X.~Li, L.~Wang, Y.~Xin, Y.~Yang, Q.~Tang, and Y.~Chen, ``Automated software
  vulnerability detection based on hybrid neural network,'' \emph{Applied
  Sciences}, vol.~11, no.~7, p. 3201, 2021.

\bibitem{mirskyvulchecker}
Y.~Mirsky, G.~Macon, M.~Brown, C.~Yagemann, M.~Pruett, E.~Downing,
  S.~Mertoguno, and W.~Lee, ``Vulchecker: Graph-based vulnerability
  localization in source code.''

\bibitem{209211}
P.~Black, ``\BIBforeignlanguage{en}{Sard: A software assurance reference
  dataset},'' 1970.

\bibitem{cwe}
``{CWE - Common Vulnerability Enumeration},'' \url{https://cwe.mitre.org/},
  2023.

\bibitem{olson2008advanced}
D.~L. Olson and D.~Delen, \emph{Advanced data mining techniques}.\hskip 1em
  plus 0.5em minus 0.4em\relax Springer Science \& Business Media, 2008.

\bibitem{powers2020evaluation}
D.~M. Powers, ``Evaluation: from precision, recall and f-measure to roc,
  informedness, markedness and correlation,'' \emph{arXiv preprint
  arXiv:2010.16061}, 2020.

\bibitem{xu2020automatic}
Z.~Xu, Y.~Zhang, L.~Zheng, L.~Xia, C.~Bao, Z.~Wang, and Y.~Liu, ``Automatic hot
  patch generation for android kernels,'' in \emph{Proceedings of the 29th
  USENIX Conference on Security Symposium}, 2020, pp. 2397--2414.

\bibitem{christou2022ivysyn}
N.~Christou, D.~Jin, V.~Atlidakis, B.~Ray, and V.~P. Kemerlis, ``Ivysyn:
  Automated vulnerability discovery for deep learning frameworks,'' \emph{arXiv
  preprint arXiv:2209.14921}, 2022.

\bibitem{mulliner2013patchdroid}
C.~Mulliner, J.~Oberheide, W.~Robertson, and E.~Kirda, ``Patchdroid: Scalable
  third-party security patches for android devices,'' in \emph{Proceedings of
  the 29th Annual Computer Security Applications Conference}, 2013, pp.
  259--268.

\bibitem{arnold2009ksplice}
J.~Arnold and M.~F. Kaashoek, ``Ksplice: Automatic rebootless kernel updates,''
  in \emph{Proceedings of the 4th ACM European conference on Computer systems},
  2009, pp. 187--198.

\bibitem{duan2019automating}
R.~Duan, A.~Bijlani, Y.~Ji, O.~Alrawi, Y.~Xiong, M.~Ike, B.~Saltaformaggio, and
  W.~Lee, ``Automating patching of vulnerable open-source software versions in
  application binaries.'' in \emph{NDSS}, 2019.

\bibitem{black2018juliet}
P.~E. Black and P.~E. Black, \emph{Juliet 1.3 test suite: Changes from
  1.2}.\hskip 1em plus 0.5em minus 0.4em\relax US Department of Commerce,
  National Institute of Standards and Technology, 2018.

\bibitem{openssl}
{The Open{SSL} Project}, ``{OpenSSL}: The open source toolkit for {SSL/TLS},''
  April 2003, \url{www.openssl.org}.

\bibitem{ghanbari2019practical}
A.~Ghanbari, S.~Benton, and L.~Zhang, ``Practical program repair via bytecode
  mutation,'' in \emph{Proceedings of the 28th ACM SIGSOFT International
  Symposium on Software Testing and Analysis}, 2019, pp. 19--30.

\bibitem{hua2018sketchfix}
J.~Hua, M.~Zhang, K.~Wang, and S.~Khurshid, ``Sketchfix: a tool for automated
  program repair approach using lazy candidate generation,'' in
  \emph{Proceedings of the 2018 26th ACM Joint Meeting on European Software
  Engineering Conference and Symposium on the Foundations of Software
  Engineering}, 2018, pp. 888--891.

\bibitem{liu2019tbar}
K.~Liu, A.~Koyuncu, D.~Kim, and T.~F. Bissyand{\'e}, ``Tbar: Revisiting
  template-based automated program repair,'' in \emph{Proceedings of the 28th
  ACM SIGSOFT International Symposium on Software Testing and Analysis}, 2019,
  pp. 31--42.

\bibitem{martinez2016astor}
M.~Martinez and M.~Monperrus, ``Astor: A program repair library for java,'' in
  \emph{Proceedings of the 25th International Symposium on Software Testing and
  Analysis}, 2016, pp. 441--444.

\bibitem{long2015staged}
F.~Long and M.~Rinard, ``Staged program repair with condition synthesis,'' in
  \emph{Proceedings of the 2015 10th Joint Meeting on Foundations of Software
  Engineering}, 2015, pp. 166--178.

\bibitem{demarco2014automatic}
F.~DeMarco, J.~Xuan, D.~Le~Berre, and M.~Monperrus, ``Automatic repair of buggy
  if conditions and missing preconditions with smt,'' in \emph{Proceedings of
  the 6th international workshop on constraints in software testing,
  verification, and analysis}, 2014, pp. 30--39.

\bibitem{chen2019sequencer}
Z.~Chen, S.~Kommrusch, M.~Tufano, L.-N. Pouchet, D.~Poshyvanyk, and
  M.~Monperrus, ``Sequencer: Sequence-to-sequence learning for end-to-end
  program repair,'' \emph{IEEE Transactions on Software Engineering}, vol.~47,
  no.~9, pp. 1943--1959, 2019.

\bibitem{li2020dlfix}
Y.~Li, S.~Wang, and T.~N. Nguyen, ``Dlfix: Context-based code transformation
  learning for automated program repair,'' in \emph{Proceedings of the ACM/IEEE
  42nd International Conference on Software Engineering}, 2020, pp. 602--614.

\bibitem{ye2022neural}
H.~Ye, M.~Martinez, and M.~Monperrus, ``Neural program repair with
  execution-based backpropagation,'' in \emph{Proceedings of the 44th
  International Conference on Software Engineering}, 2022, pp. 1506--1518.

\bibitem{zhu2021syntax}
Q.~Zhu, Z.~Sun, Y.-a. Xiao, W.~Zhang, K.~Yuan, Y.~Xiong, and L.~Zhang, ``A
  syntax-guided edit decoder for neural program repair,'' in \emph{Proceedings
  of the 29th ACM Joint Meeting on European Software Engineering Conference and
  Symposium on the Foundations of Software Engineering}, 2021, pp. 341--353.

\bibitem{lin2017quixbugs}
D.~Lin, J.~Koppel, A.~Chen, and A.~Solar-Lezama, ``Quixbugs: A multi-lingual
  program repair benchmark set based on the quixey challenge,'' in
  \emph{Proceedings Companion of the 2017 ACM SIGPLAN international conference
  on systems, programming, languages, and applications: software for humanity},
  2017, pp. 55--56.

\bibitem{monperrus2018automatic}
M.~Monperrus, ``Automatic software repair: a bibliography,'' \emph{ACM
  Computing Surveys (CSUR)}, vol.~51, no.~1, pp. 1--24, 2018.

\bibitem{shi2022backporting}
Y.~Shi, Y.~Zhang, T.~Luo, X.~Mao, Y.~Cao, Z.~Wang, Y.~Zhao, Z.~Huang, and
  M.~Yang, ``Backporting security patches of web applications: A prototype
  design and implementation on injection vulnerability patches,'' in \emph{31st
  USENIX Security Symposium (USENIX Security 22)}, 2022, pp. 1993--2010.

\bibitem{sidiroglou2005countering}
S.~Sidiroglou and A.~D. Keromytis, ``Countering network worms through automatic
  patch generation,'' \emph{IEEE Security \& Privacy}, vol.~3, no.~6, pp.
  41--49, 2005.

\bibitem{martinez2015mining}
M.~Martinez and M.~Monperrus, ``Mining software repair models for reasoning on
  the search space of automated program fixing,'' \emph{Empirical Software
  Engineering}, vol.~20, pp. 176--205, 2015.

\bibitem{kim2013automatic}
D.~Kim, J.~Nam, J.~Song, and S.~Kim, ``Automatic patch generation learned from
  human-written patches,'' in \emph{2013 35th International Conference on
  Software Engineering (ICSE)}.\hskip 1em plus 0.5em minus 0.4em\relax IEEE,
  2013, pp. 802--811.

\bibitem{ma2017vurle}
S.~Ma, F.~Thung, D.~Lo, C.~Sun, and R.~H. Deng, ``Vurle: Automatic
  vulnerability detection and repair by learning from examples,'' in
  \emph{Computer Security--ESORICS 2017: 22nd European Symposium on Research in
  Computer Security, Oslo, Norway, September 11-15, 2017, Proceedings, Part II
  22}.\hskip 1em plus 0.5em minus 0.4em\relax Springer, 2017, pp. 229--246.

\bibitem{yagemann2021arcus}
C.~Yagemann, M.~Pruett, S.~P. Chung, K.~Bittick, B.~Saltaformaggio, and W.~Lee,
  ``Arcus: Symbolic root cause analysis of exploits in production systems.'' in
  \emph{USENIX Security Symposium}, 2021, pp. 1989--2006.

\bibitem{musuvathi2008finding}
M.~Musuvathi, S.~Qadeer, T.~Ball, G.~Basler, P.~A. Nainar, and I.~Neamtiu,
  ``Finding and reproducing heisenbugs in concurrent programs.'' in
  \emph{OSDI}, vol.~8, no. 2008, 2008.

\bibitem{ahmed2014impact}
I.~Ahmed, N.~Mohan, and C.~Jensen, ``The impact of automatic crash reports on
  bug triaging and development in mozilla,'' in \emph{Proceedings of The
  International Symposium on Open Collaboration}, 2014, pp. 1--8.

\bibitem{blazytko2020aurora}
T.~Blazytko, M.~Schl{\"o}gel, C.~Aschermann, A.~Abbasi, J.~Frank,
  S.~W{\"o}rner, and T.~Holz, ``Aurora: Statistical crash analysis for
  automated root cause explanation,'' in \emph{Proceedings of the 29th USENIX
  Conference on Security Symposium}, 2020, pp. 235--252.

\bibitem{attariyan2012x}
M.~Attariyan, M.~Chow, and J.~Flinn, ``X-ray: Automating root-cause diagnosis
  of performance anomalies in production software,'' in \emph{Presented as part
  of the 10th $\{$USENIX$\}$ Symposium on Operating Systems Design and
  Implementation ($\{$OSDI$\}$ 12)}, 2012, pp. 307--320.

\bibitem{cui2018rept}
W.~Cui, X.~Ge, B.~Kasikci, B.~Niu, U.~Sharma, R.~Wang, and I.~Yun,
  ``$\{$REPT$\}$: Reverse debugging of failures in deployed software,'' in
  \emph{13th $\{$USENIX$\}$ Symposium on Operating Systems Design and
  Implementation ($\{$OSDI$\}$ 18)}, 2018, pp. 17--32.

\bibitem{cui2016retracer}
W.~Cui, M.~Peinado, S.~K. Cha, Y.~Fratantonio, and V.~P. Kemerlis, ``Retracer:
  Triaging crashes by reverse execution from partial memory dumps,'' in
  \emph{Proceedings of the 38th International Conference on Software
  Engineering}, 2016, pp. 820--831.

\bibitem{xu2017postmortem}
J.~Xu, D.~Mu, X.~Xing, P.~Liu, P.~Chen, and B.~Mao, ``Postmortem program
  analysis with hardware-enhanced post-crash artifacts.'' in \emph{USENIX
  Security Symposium}, 2017, pp. 17--32.

\bibitem{reiter2022automatically}
P.~Reiter, H.~J. Tay, W.~Weimer, A.~Doup{\'e}, R.~Wang, and S.~Forrest,
  ``Automatically mitigating vulnerabilities in x86 binary programs via
  partially recompilable decompilation,'' \emph{arXiv preprint
  arXiv:2202.12336}, 2022.

\bibitem{mantovani2022convergence}
A.~Mantovani, L.~Compagna, Y.~Shoshitaishvili, and D.~Balzarotti, ``The
  convergence of source code and binary vulnerability discovery--a case
  study,'' in \emph{Proceedings of the 2022 ACM on Asia Conference on Computer
  and Communications Security}, 2022, pp. 602--615.

\bibitem{verbeek2020sound}
F.~Verbeek, P.~Olivier, and B.~Ravindran, ``Sound c code decompilation for a
  subset of x86-64 binaries,'' in \emph{Software Engineering and Formal
  Methods: 18th International Conference, SEFM 2020, Amsterdam, The
  Netherlands, September 14--18, 2020, Proceedings 18}.\hskip 1em plus 0.5em
  minus 0.4em\relax Springer, 2020, pp. 247--264.

\bibitem{schwartz2013native}
E.~J. Schwartz, J.~Lee, M.~Woo, and D.~Brumley, ``Native x86 decompilation
  using semantics-preserving structural analysis and iterative control-flow
  structuring,'' in \emph{Proceedings of the USENIX Security Symposium},
  vol.~16, 2013.

\bibitem{branco2012scientific}
R.~R. Branco, G.~N. Barbosa, and P.~D. Neto, ``Scientific but not academical
  overview of malware anti-debugging, anti-disassembly and anti-vm
  technologies,'' \emph{Black Hat}, vol.~1, no. 2012, pp. 1--27, 2012.

\bibitem{schulte2018evolving}
E.~Schulte, J.~Ruchti, M.~Noonan, D.~Ciarletta, and A.~Loginov, ``Evolving
  exact decompilation,'' in \emph{Workshop on Binary Analysis Research (BAR)},
  2018.

\bibitem{ida}
\BIBentryALTinterwordspacing
Hex-Rays, ``Hex-rays,'' 2022. [Online]. Available:
  \url{https://www.hex-rays.com/decompiler/}
\BIBentrySTDinterwordspacing

\bibitem{leetcode}
\BIBentryALTinterwordspacing
``Leetcode,'' 2023. [Online]. Available: \url{https://leetcode.com/}
\BIBentrySTDinterwordspacing

\bibitem{lacomis2019dire}
J.~Lacomis, P.~Yin, E.~Schwartz, M.~Allamanis, C.~Le~Goues, G.~Neubig, and
  B.~Vasilescu, ``Dire: A neural approach to decompiled identifier naming,'' in
  \emph{2019 34th IEEE/ACM International Conference on Automated Software
  Engineering (ASE)}.\hskip 1em plus 0.5em minus 0.4em\relax IEEE, 2019, pp.
  628--639.

\bibitem{chen2022augmenting}
Q.~Chen, J.~Lacomis, E.~J. Schwartz, C.~Le~Goues, G.~Neubig, and B.~Vasilescu,
  ``Augmenting decompiler output with learned variable names and types,'' in
  \emph{31st USENIX Security Symposium (USENIX Security 22)}, 2022, pp.
  4327--4343.

\bibitem{llvm}
\BIBentryALTinterwordspacing
``Llvm commandline guide.'' 2023. [Online]. Available:
  \url{https://llvm.org/docs/CommandGuide/llvm-cov.html}
\BIBentrySTDinterwordspacing

\bibitem{SQLite}
\BIBentryALTinterwordspacing
``Sqlite,'' 2023. [Online]. Available: \url{https://www.sqlite.org/}
\BIBentrySTDinterwordspacing

\bibitem{EGLIBC}
\BIBentryALTinterwordspacing
``Eglibc,'' 2023. [Online]. Available: \url{http://www.eglibc.org/home}
\BIBentrySTDinterwordspacing

\bibitem{Toybox}
\BIBentryALTinterwordspacing
``Toybox,'' 2023. [Online]. Available: \url{https://landley.net/toybox}
\BIBentrySTDinterwordspacing

\bibitem{BusyBox}
\BIBentryALTinterwordspacing
``Busybox,'' 2023. [Online]. Available: \url{https://busybox.net}
\BIBentrySTDinterwordspacing

\bibitem{jiang2016jred}
Y.~Jiang, D.~Wu, and P.~Liu, ``Jred: Program customization and bloatware
  mitigation based on static analysis,'' in \emph{2016 IEEE 40th annual
  computer software and applications conference (COMPSAC)}, vol.~1.\hskip 1em
  plus 0.5em minus 0.4em\relax IEEE, 2016, pp. 12--21.

\bibitem{quach2018debloating}
A.~Quach, A.~Prakash, and L.~Yan, ``Debloating software through piece-wise
  compilation and loading,'' in \emph{27th $\{$USENIX$\}$ Security Symposium
  ($\{$USENIX$\}$ Security 18)}, 2018, pp. 869--886.

\bibitem{rastogi2017cimplifier}
V.~Rastogi, D.~Davidson, L.~De~Carli, S.~Jha, and P.~McDaniel, ``Cimplifier:
  automatically debloating containers,'' in \emph{Proceedings of the 2017 11th
  Joint Meeting on Foundations of Software Engineering}, 2017, pp. 476--486.

\bibitem{ghavamnia2020temporal}
S.~Ghavamnia, T.~Palit, S.~Mishra, and M.~Polychronakis, ``Temporal system call
  specialization for attack surface reduction,'' in \emph{Proceedings of the
  29th USENIX Conference on Security Symposium}, 2020, pp. 1749--1766.

\bibitem{azad2019less}
B.~A. Azad, P.~Laperdrix, and N.~Nikiforakis, ``Less is more: Quantifying the
  security benefits of debloating web applications.'' in \emph{USENIX Security
  Symposium}, 2019, pp. 1697--1714.

\bibitem{heo2018effective}
K.~Heo, W.~Lee, P.~Pashakhanloo, and M.~Naik, ``Effective program debloating
  via reinforcement learning,'' in \emph{Proceedings of the 2018 ACM SIGSAC
  Conference on Computer and Communications Security}, 2018, pp. 380--394.

\bibitem{ghaffarinia2019binary}
M.~Ghaffarinia and K.~W. Hamlen, ``Binary control-flow trimming,'' in
  \emph{Proceedings of the 2019 ACM SIGSAC Conference on Computer and
  Communications Security}, 2019, pp. 1009--1022.

\bibitem{xue2019hecate}
H.~Xue, Y.~Chen, G.~Venkataramani, and T.~Lan, ``Hecate: Automated
  customization of program and communication features to reduce attack
  surfaces,'' in \emph{Security and Privacy in Communication Networks: 15th EAI
  International Conference, SecureComm 2019, Orlando, FL, USA, October 23--25,
  2019, Proceedings, Part II 15}.\hskip 1em plus 0.5em minus 0.4em\relax
  Springer, 2019, pp. 305--319.

\bibitem{bzip2}
\BIBentryALTinterwordspacing
``bzip2 - home,'' 2023. [Online]. Available:
  \url{https://sourceware.org/bzip2/}
\BIBentrySTDinterwordspacing

\bibitem{zhou2022ferry}
S.~Zhou, Z.~Yang, D.~Qiao, P.~Liu, M.~Yang, Z.~Wang, and C.~Wu,
  ``Ferry:$\{$State-Aware$\}$ symbolic execution for exploring
  $\{$State-Dependent$\}$ program paths,'' in \emph{31st USENIX Security
  Symposium (USENIX Security 22)}, 2022, pp. 4365--4382.

\bibitem{boyer1975select}
R.~S. Boyer, B.~Elspas, and K.~N. Levitt, ``Select—a formal system for
  testing and debugging programs by symbolic execution,'' \emph{ACM SigPlan
  Notices}, vol.~10, no.~6, pp. 234--245, 1975.

\bibitem{howden1977symbolic}
W.~E. Howden, ``Symbolic testing and the dissect symbolic evaluation system,''
  \emph{IEEE Transactions on Software Engineering}, no.~4, pp. 266--278, 1977.

\bibitem{king1975new}
J.~C. King, ``A new approach to program testing,'' \emph{ACM Sigplan Notices},
  vol.~10, no.~6, pp. 228--233, 1975.

\bibitem{cadar2008klee}
C.~Cadar, D.~Dunbar, D.~R. Engler \emph{et~al.}, ``Klee: unassisted and
  automatic generation of high-coverage tests for complex systems programs.''
  in \emph{OSDI}, vol.~8, 2008, pp. 209--224.

\bibitem{aschermann2019redqueen}
C.~Aschermann, S.~Schumilo, T.~Blazytko, R.~Gawlik, and T.~Holz, ``Redqueen:
  Fuzzing with input-to-state correspondence.'' in \emph{NDSS}, vol.~19, 2019,
  pp. 1--15.

\bibitem{chen2018angora}
P.~Chen and H.~Chen, ``Angora: Efficient fuzzing by principled search,'' in
  \emph{2018 IEEE Symposium on Security and Privacy (SP)}.\hskip 1em plus 0.5em
  minus 0.4em\relax IEEE, 2018, pp. 711--725.

\bibitem{chen2019matryoshka}
P.~Chen, J.~Liu, and H.~Chen, ``Matryoshka: fuzzing deeply nested branches,''
  in \emph{Proceedings of the 2019 ACM SIGSAC Conference on Computer and
  Communications Security}, 2019, pp. 499--513.

\bibitem{chipounov2011s2e}
V.~Chipounov, V.~Kuznetsov, and G.~Candea, ``S2e: A platform for in-vivo
  multi-path analysis of software systems,'' \emph{Acm Sigplan Notices},
  vol.~46, no.~3, pp. 265--278, 2011.

\bibitem{anand2008demand}
S.~Anand, P.~Godefroid, and N.~Tillmann, ``Demand-driven compositional symbolic
  execution,'' in \emph{Tools and Algorithms for the Construction and Analysis
  of Systems: 14th International Conference, TACAS 2008, Held as Part of the
  Joint European Conferences on Theory and Practice of Software, ETAPS 2008,
  Budapest, Hungary, March 29-April 6, 2008. Proceedings 14}.\hskip 1em plus
  0.5em minus 0.4em\relax Springer, 2008, pp. 367--381.

\bibitem{schwartz2015conflict}
D.~Schwartz-Narbonne, M.~Sch{\"a}f, D.~Jovanovi{\'c}, P.~R{\"u}mmer, and
  T.~Wies, ``Conflict-directed graph coverage,'' in \emph{NASA Formal Methods:
  7th International Symposium, NFM 2015, Pasadena, CA, USA, April 27-29, 2015,
  Proceedings 7}.\hskip 1em plus 0.5em minus 0.4em\relax Springer, 2015, pp.
  327--342.

\bibitem{ma2011directed}
K.-K. Ma, K.~Yit~Phang, J.~S. Foster, and M.~Hicks, ``Directed symbolic
  execution,'' in \emph{Static Analysis: 18th International Symposium, SAS
  2011, Venice, Italy, September 14-16, 2011. Proceedings 18}.\hskip 1em plus
  0.5em minus 0.4em\relax Springer, 2011, pp. 95--111.

\bibitem{kuznetsov2012efficient}
V.~Kuznetsov, J.~Kinder, S.~Bucur, and G.~Candea, ``Efficient state merging in
  symbolic execution,'' \emph{Acm Sigplan Notices}, vol.~47, no.~6, pp.
  193--204, 2012.

\bibitem{poeplau2020symbolic}
S.~Poeplau and A.~Francillon, ``Symbolic execution with symcc: Don't interpret,
  compile!'' in \emph{Proceedings of the 29th USENIX Conference on Security
  Symposium}, 2020, pp. 181--198.

\bibitem{zhu2022fuzzing}
X.~Zhu, S.~Wen, S.~Camtepe, and Y.~Xiang, ``Fuzzing: a survey for roadmap,''
  \emph{ACM Computing Surveys (CSUR)}, vol.~54, no. 11s, pp. 1--36, 2022.

\bibitem{manes2019art}
V.~J. Man{\`e}s, H.~Han, C.~Han, S.~K. Cha, M.~Egele, E.~J. Schwartz, and
  M.~Woo, ``The art, science, and engineering of fuzzing: A survey,''
  \emph{IEEE Transactions on Software Engineering}, vol.~47, no.~11, pp.
  2312--2331, 2019.

\bibitem{peng2018t}
H.~Peng, Y.~Shoshitaishvili, and M.~Payer, ``T-fuzz: fuzzing by program
  transformation,'' in \emph{2018 IEEE Symposium on Security and Privacy
  (SP)}.\hskip 1em plus 0.5em minus 0.4em\relax IEEE, 2018, pp. 697--710.

\bibitem{fayaz2016buzz}
S.~K. Fayaz, T.~Yu, Y.~Tobioka, S.~Chaki, and V.~Sekar, ``$\{$BUZZ$\}$: Testing
  context-dependent policies in stateful networks,'' in \emph{13th
  $\{$USENIX$\}$ Symposium on Networked Systems Design and Implementation
  ($\{$NSDI$\}$ 16)}, 2016, pp. 275--289.

\bibitem{rawat2017vuzzer}
S.~Rawat, V.~Jain, A.~Kumar, L.~Cojocar, C.~Giuffrida, and H.~Bos, ``Vuzzer:
  Application-aware evolutionary fuzzing.'' in \emph{NDSS}, vol.~17, 2017, pp.
  1--14.

\bibitem{kaksonen2001software}
R.~Kaksonen, M.~Laakso, and A.~Takanen, ``Software security assessment through
  specification mutations and fault injection,'' in \emph{Communications and
  Multimedia Security Issues of the New Century: IFIP TC6/TC11 Fifth Joint
  Working Conference on Communications and Multimedia Security (CMS’01) May
  21--22, 2001, Darmstadt, Germany}.\hskip 1em plus 0.5em minus 0.4em\relax
  Springer, 2001, pp. 173--183.

\bibitem{dharma}
``{dharma},'' \url{https://github.com/MozillaSecurity/dharma}.

\bibitem{vuagnoux2005autodafe}
M.~Vuagnoux, ``Autodafe: An act of software torture,'' in \emph{Proceedings of
  the 22th Chaos Communication Congress}, no. CONF.\hskip 1em plus 0.5em minus
  0.4em\relax Chaos Computer Club, 2005, pp. 47--58.

\bibitem{abdelnur2007kif}
H.~J. Abdelnur, R.~State, and O.~Festor, ``Kif: a stateful sip fuzzer,'' in
  \emph{Proceedings of the 1st international Conference on Principles, Systems
  and Applications of IP Telecommunications}, 2007, pp. 47--56.

\bibitem{stephens2016driller}
N.~Stephens, J.~Grosen, C.~Salls, A.~Dutcher, R.~Wang, J.~Corbetta,
  Y.~Shoshitaishvili, C.~Kruegel, and G.~Vigna, ``Driller: Augmenting fuzzing
  through selective symbolic execution.'' in \emph{NDSS}, vol.~16, no. 2016,
  2016, pp. 1--16.

\bibitem{yun2018qsym}
I.~Yun, S.~Lee, M.~Xu, Y.~Jang, and T.~Kim, ``$\{$QSYM$\}$: A practical
  concolic execution engine tailored for hybrid fuzzing,'' in \emph{27th
  $\{$USENIX$\}$ Security Symposium ($\{$USENIX$\}$ Security 18)}, 2018, pp.
  745--761.

\bibitem{afl}
\BIBentryALTinterwordspacing
``american fuzzy lop,'' 2023. [Online]. Available:
  \url{https://lcamtuf.coredump.cx/afl/}
\BIBentrySTDinterwordspacing

\bibitem{afl++}
A.~Fioraldi, D.~Maier, H.~Ei{\ss}feldt, and M.~Heuse, ``Afl++ combining
  incremental steps of fuzzing research,'' in \emph{Proceedings of the 14th
  USENIX Conference on Offensive Technologies}, 2020, pp. 10--10.

\bibitem{libfuzzer}
\BIBentryALTinterwordspacing
``libfuzzer – a library for coverage-guided fuzz testing,'' 2023. [Online].
  Available: \url{https://llvm.org/docs/LibFuzzer.html}
\BIBentrySTDinterwordspacing

\bibitem{serebryany2017oss}
K.~Serebryany, ``Oss-fuzz-google’s continuous fuzzing service for open source
  software,'' in \emph{USENIX Security symposium}.\hskip 1em plus 0.5em minus
  0.4em\relax USENIX Association, 2017.

\bibitem{metzman2021fuzzbench}
J.~Metzman, L.~Szekeres, L.~Simon, R.~Sprabery, and A.~Arya, ``Fuzzbench: an
  open fuzzer benchmarking platform and service,'' in \emph{Proceedings of the
  29th ACM joint meeting on European software engineering conference and
  symposium on the foundations of software engineering}, 2021, pp. 1393--1403.

\bibitem{collobert2002torch}
R.~Collobert, S.~Bengio, and J.~Mari{\'e}thoz, ``Torch: a modular machine
  learning software library,'' Idiap, Tech. Rep., 2002.

\bibitem{abadi2016tensorflow}
M.~Abadi, ``Tensorflow: learning functions at scale,'' in \emph{Proceedings of
  the 21st ACM SIGPLAN International Conference on Functional Programming},
  2016, pp. 1--1.

\end{thebibliography}

\end{document}